\documentclass{article}

\input{dynlift.sty}

\begin{document}

\title{Proto-Quipper with dynamic lifting}
\author{Peng Fu$^{1}$, Kohei Kishida$^{2}$, Neil J. Ross$^{1}$, Peter Selinger$^{1}$\\
  \small $^{1}$Dalhousie University \\
  \small $^{2}$University of Illinois Urbana-Champaign
}
\date{}
\maketitle

\begin{abstract}
Quipper is a functional programming language for quantum computing.
Proto-Quipper is a family of languages aiming to provide a formal
foundation for Quipper. In this paper, we extend Proto-Quipper-M with
a construct called \emph{dynamic lifting}, which is present in Quipper.  By
virtue of being a circuit description language, Proto-Quipper has two
separate runtimes: circuit generation time and circuit execution
time. Values that are known at circuit generation time are called
\emph{parameters}, and values that are known at circuit execution time
are called \emph{states}. Dynamic lifting is an operation that enables a state,
such as the result of a measurement, to be lifted to a parameter,
where it can influence the generation of the next portion of the
circuit. As a result, dynamic lifting enables Proto-Quipper programs
to interleave classical and quantum computation. We describe the
syntax of a language we call Proto-Quipper-Dyn. Its type system uses a
system of modalities to keep track of the use of dynamic lifting. We
also provide an operational semantics, as well as an abstract
categorical semantics for dynamic lifting based on enriched category
theory. We prove that both the type system and the operational
semantics are sound with respect to our categorical
semantics. Finally, we give some examples of Proto-Quipper-Dyn
programs that make essential use of dynamic lifting.
\end{abstract}

\section{Introduction}
\subsection{Quipper and Proto-Quipper}
Quipper is a functional programming language for quantum computing
\cite{GLRSV2013-pldi,GLRSV2013-rc}. The overall aim of Quipper is to
allow quantum algorithms to be specified at a level of abstraction
that is similar to how the algorithm might be described in a research
paper, and to compile this down to the level of individual quantum
gates, producing a logical quantum circuit. Quipper has been used to
program a set of nontrivial algorithms from the quantum computing
literature, and it has been used to generate quantum circuits
consisting of trillions of gates. As a circuit description language,
Quipper shares some of the traits of hardware description
languages. In particular, it has two notions of runtime: The first of
these is \emph{circuit generation time}. This is when a Quipper
program is run to generate a quantum circuit. The second is
\emph{circuit execution time}. This is when a quantum circuit is
executed by a quantum computer or a simulator. 

Quipper is a practical language, implemented as an embedded language
in Haskell. As such, it lacks formal foundations such as operational
and denotational semantics. This motivates the development of
Proto-Quipper, a family of experimental languages that aim to provide
formal semantics for fragments of Quipper.  Proto-Quipper-S features a
linear type system with subtyping as well as an operational semantics
\cite{ross2015algebraic}. Proto-Quipper-M has a linear type system
without subtyping, but with a sound categorical semantics in addition
to its operation semantics \cite{RS2017-pqmodel}. More recently,
Proto-Quipper-D was introduced, which features a type system with
linear dependent types as well as a fibrational categorical semantics
\cite{FKS2020-lindep,FKRS2020-dpq-tutorial}.

\subsection{Dynamic lifting and the interaction of the two runtimes}
Proto-Quipper, like Quipper, distinguishes two runtimes.  Moreover,
Proto-Quipper gives a formal account of parameters and states.  A
\emph{parameter} is a value that is known at circuit generation time,
such as a boolean value for an if-then-else expression. A \emph{state}
is a value that is only known at circuit execution time, such as the
actual state of a qubit or classical bit in a circuit. The type system
of Proto-Quipper reflects this distinction. Among the types, there is
a subset of \emph{parameter types}, such as $\Nat$ and $\Bool$, whose
elements can be duplicated and discarded. There is also a subset of
\emph{state types}, such as $\Qubit$ and $\Bit$, which are
linear so that their elements cannot in general be duplicated or
discarded. One of the fundamental design decisions of Proto-Quipper is
that parameter types and
state types belong to the same universe of types, so that one can form
compound types that are part parameter and part state. An example of
this is the type $\Bool\otimes\Qubit$, whose elements are pairs of a
boolean (a parameter) and a qubit (a state). Another example is the
type of lists of qubits. Here,
the length of the list is a parameter
(known at circuit generation time), but the actual qubits in the list
are states (known at circuit execution time). In this way, Proto-Quipper
differs, e.g., from QWire, an embedded quantum circuit description
language in which parameters
and states belong to separate universes {\cite{paykin2017qwire}}.

In Quipper, the two runtimes can interact with each other. A priori,
it is clear that states can depend on parameters. For example, we can
initialize a qubit based on a boolean parameter, simply by inserting a
gate at circuit generation time to initialize the qubit in one state
or another. The opposite direction is more complicated. Usually,
circuit execution happens \emph{after} circuit generation, and in this
case, it is clear that a state cannot be converted to a
parameter. However, there are some quantum algorithms that require
circuit generation and circuit execution to be interleaved. Here, a
state, such as the outcome of a measurement in a circuit, may be used
to inform the generation of the next part of the circuit. To enable
such interleaving, Quipper provides a construct called
\emph{dynamic lifting}, which enables a state to be lifted to a
parameter in certain situations. For example, dynamic lifting permits
the result of a measurement, which is a state of type $\Bit$, to be
lifted to a parameter of type $\Bool$. It is important to note that
dynamic lifting is an \emph{expensive} operation, as it requires
control to pass from circuit evaluation time back to circuit
generation time. This requires the real-time quantum computer to put
all of its active qubits into long-term storage while spending an
indeterminate amount of time awaiting further instructions from the
classical computer in charge of circuit generation.

Dynamic lifting is important because it can be used to express quantum
algorithms that require interleaving circuit execution time and
circuit generation time. While there are many quantum algorithms that
do not require such interleaving, there are some that do. An example
is \textit{magic state distillation} \cite{bravyi2005universal}. Here,
the goal is to prepare a qubit in some target state. We start with a
large number of qubits, say $n$ of them, each of which is a rough
approximation of the target state. We then apply a probabilistic
``distillation'' procedure which yields on average, say, $n/4$
qubits that
are better approximations of the target state; the remaining qubits
are wasted. By repeated distillation steps, we eventually wind up with
a small number of qubits that are excellent approximations of the
target state. In such a situation, dynamic lifting is essential
because after each distillation step, we must throw away the wasted
qubits, but we do not know ahead of time which ones (or indeed, how
many) there will be. Thus which future gates will be applied depends
on the outcomes of previous measurements. With the help of dynamic
lifting, these algorithms can be naturally expressed as functions in
the programming language.

The concept of dynamic lifting is different from measurement, and the
two should not be confused. Measurement is merely a gate in a circuit,
turning a quantum bit (a state) into a classical bit (also a
state). Dynamic lifting is an operation of the programming language,
turning a classical bit (a state) into a boolean (a parameter).

\subsection{A type system for dynamic lifting}
Previous versions of Proto-Quipper lacked dynamic lifting. Modeling
dynamic lifting is a challenging problem. To better understand the
issues involved, it is useful to know that there are two things that
can be done with circuits in Proto-Quipper. On the one hand, circuits
can be run on a quantum device. On the other hand, circuits can be
\emph{boxed}. A boxed circuit is a data structure that contains a
circuit that has \emph{already been generated}, i.e., an actual list
of gates, rather than merely instructions for how to generate such a
list. As a result, a boxed circuit can be used as a building block for
all kinds of things. In the simplest case, it can be re-used in the
construction of other circuits. But it can also be inspected and
manipulated in other ways, such as by applying gate transformations
(systematically replacing gates by other gates), by adding things like
error correction, or by rewriting the circuit to simplify it, among
many other possibilities. Boxed circuits can also be reversed, which is
used in many quantum algorithms, for example to uncompute ancillas.
The ability to box circuits is crucial to Quipper's ability to express
algorithms at a natural level of abstraction, because algorithms are
often described in terms of meta-operations on circuits.

Now it is clear that dynamic lifting only makes sense in the context
of a circuit that is actually being executed, rather than one that is
merely being boxed. We will keep track of this in the programming
language by adding a modality to the type system and a corresponding
monad to the semantics. The modality should be thought of as denoting
``boxability''. For example, a function of type $\Qubit
\multimap_1\Qubit$ represents a circuit that can be boxed or executed,
i.e., that does not use dynamic lifting, whereas a function of type
$\Qubit\multimap_0\Qubit$ represents a quantum operation that can only
be executed but not boxed.

Before we can describe the operational or denotational semantics of
Proto-Quipper-Dyn, we must be more precise about what we mean by a
``circuit''. We must also specify what it means to ``execute'' a
circuit. There are many different notions of circuits, differing, for
example, in which collection of gates is provided. Rather than
specializing to one of these, we take a more general point of view: a
\emph{circuit} is simply a morphism in a small symmetric monoidal category
$\m$, which we assume to be given ahead of time, but otherwise
arbitrary (subject to some properties). Similarly, for the execution
of circuits, we assume given another small symmetric monoidal category $\q$
of \emph{quantum operations}. Conceptually, the morphisms of $\m$ are
\emph{syntactic} entities; thus, $\m$ is typically a category that is
free generated (say by a collection of gates). On the other hand, we
think of the morphisms of $\q$ as \emph{physical} operations, which
can be performed on a quantum computer. The categories $\m$ and $\q$
have the same objects, and there is a symmetric monoidal \textit{interpretation functor}
$J : \m \to \q$.

Operationally, dynamic lifting is an operation that reads the state of
a bit in $\q$, and returns a boolean value. Since a bit state can be
the result of a measurement, the read operation for dynamic lifting is
nondeterministic, i.e., it can return different boolean values with
probabilities governed by measurements. The nondeterministic nature of
the dynamic lifting suggests that it should be modeled as a monadic
operation \cite{moggi1991notions}.

We therefore conceptualize the types of Proto-Quipper-Dyn as
the objects of a single category $\A$, with a monad $T:\A\to\A$,
called the \emph{dynamic lifting monad}. This will be done in such a
way that $\m$ is fully embedded in $\A$, and $\q$ is fully embedded in
the Kleisli category $\Kl_{T}(\A)$, in a way that makes the following
diagram commute.
\[
  \begin{tikzcd}
  \m\arrow[r, hook, "\psi"] \arrow[d, "J"] & \A \arrow[d, "E"]\\
  \q \arrow[r, hook, "\phi"] & \Kl_{T}(\A)
  \end{tikzcd}
\]
Here, $J$ is the given interpretation functor, and $E$ is the
canonical functor from $\A$ to $\Kl_{T}(\A)$. We then model dynamic
lifting as a map $\Dyn : \Bit \to T\Bool \in \Kl_{T}(\A)$ such that
the following diagram commutes.
\[
  \begin{tikzcd}
    & \Bit \arrow[d, "\Dyn"]\\
    \Bool \arrow[r, "\eta"] \arrow[ur, "\mathsf{init}"]  & T\Bool
  \end{tikzcd}
\]
Note that dynamic lifting is a morphism of the Kleisli category; this
makes sense because it is essentially a side-effecting read
operation. More generally, any computation that potentially uses
dynamic lifting will have type $A\to TB$.

As mentioned above, our type system must also distinguish quantum
circuits that are being executed from quantum circuits that are being
boxed. Naturally, since the latter may not use dynamic lifting, they
are maps in the category $\A$ while the former are maps in the Kleisli
category $\Kl_{T}(\A)$. As a practical matter for programmer
convenience, it would be awkward to have $T$ as an explicit type
constructor that must be mentioned everywhere in the program. Instead,
we use a system of modalities to keep track of the dynamic lifting
monad $T$.  More specifically, we annotate a typing judgment with a
modality, i.e., $\Gamma \vdash_{\alpha} M : A$, where
$\alpha \in \{0, 1\}$.  When $\alpha = 0$, it means that the term $M$
represents a morphism $\interp{\Gamma} \to T \interp{A}$ in the
Kleisli category $\Kl_{T}(\A)$.  When $\alpha = 1$, it means that the
term $M$ represents a morphism $\interp{\Gamma} \to \interp{A}$ in
$\A$.  An example of the typing rule for dynamic lifting is the
following (where $\Meas : \Qubit \to \Bit$ represents the measurement
gate).
\[
  \infer{\ell : \Qubit \vdash_0 \Dyn(\Meas (\ell))  : \Bool}
  {\ell : \Qubit \vdash_1 \Meas (\ell)  : \Bit}
\]
If we have a quantum circuit $\Qubit \to \Bit$, it can be run by a
quantum computer and the measurement result of type $\Bit$ will be
lifted to a parameter of type $\Bool$. Note that the $\Dyn$ operation
sets the modality of the typing judgment to $0$, and as a result, we
have a map $\Qubit \to T\Bool$ in the Kleisli category. The use of
modalities in our type system ensures that the term $\Meas (\ell)$ can
be turned into a boxed circuit, whereas it will be a compile time
typing error to try to box the term $\Dyn(\Meas (\ell))$.

\subsection{Operational semantics}

Next, let us take a look at the operational semantics
of Proto-Quipper-Dyn. In previous
versions of Proto-Quipper, the operational semantics used
configurations of the form $(\cC,M)$, where $\cC$ is the circuit being
currently constructed, and $M$ is a term. On the other hand, in the
quantum lambda calculus {\cite{SV2009-qlambdabook}}, which is not a
circuit construction language but intended to run directly on a
quantum computer, the operational semantics used configurations of the
form $(Q,M)$, where $Q$ is the current quantum state and $M$ is a
term.

In a sense, Proto-Quipper-Dyn is a combination of
these prior languages: it is a language for circuit construction (via
the \emph{boxing} operation), but it is also a language for running
quantum operations (as otherwise dynamic lifting would not be
possible). Consequently, our operational semantics uses \emph{both}
kinds of configurations: those of the form $(Q,M)$ are only used for
top-level computations that actually run on a quantum device, and
those of the form $(\cC,M)$ are used during boxing. These two kinds of
configurations correspond closely to the two runtimes, since
configuration of the form $(\cC,M)$ are used for circuit construction
and those of the form $(Q,M)$ are used for circuit execution. They
also correspond to the two categories $\m$ and $\q$.

Consequently, the evaluation rules take two different
forms. Evaluation at circuit generation time takes the form
$(\cC, M) \Downarrow (\cC', V)$, where $\cC$ is a circuit. The type
system ensures that such an evaluation does not involve dynamic
lifting, so it can be done entirely with a classical computer and the
evaluation is deterministic. On the other hand, evaluation at circuit
execution time takes the form
$(Q, M) \Downarrow \sum_{i}p_{i}(Q_{i}, V_{i})$, where $Q$ represents
a quantum state. Since $M$ can use dynamic lifting, the result of such
an evaluation rule is probabilistic, with outcome $(Q_{i}, V_{i})$
happening with probability $p_{i}$.

\subsection{Related work}

A common misunderstanding is that dynamic lifting means performing
measurements during the execution of a quantum program. That would not
be a new feature; indeed, the ability to perform on-the-fly
measurements was already present in the earliest quantum programming
languages, such as
{\cite{Oem98,selinger2004towards,SV2009-qlambdabook}}. Rather, dynamic
lifting is an operation that only makes sense in the context of a
{\emph{circuit description language}}, where circuits are not executed
during the circuit generation phase. Dynamic lifting is the transfer
of information from the circuit execution environment back to the
circuit generation environment. Therefore, in the following discussion
of related work, we do not include comparisons with most papers on
languages that include measurement but do not have separate circuit
generation and circuit execution times.

One of the features of the present work, and of Proto-Quipper in
general, is that it works with the standard notion of quantum circuits
{\cite{nielsen2002quantum}}, which are basically lists of gates, or
more precisely, gates that have been composed using the laws of
symmetric monoidal categories.  By contrast, some of the other notions
of dynamic lifting that appear in the literature not only add features
to the programming language, but also to the generated circuits
themselves. Relatedly, one of the features that makes boxed circuits
useful in Proto-Quipper is that they are actual data structures. Here,
by a ``data structure'', we mean data that can be queried, for example
via a case distinction or pattern matching. This is different from a
``thunk'', such as a lambda abstraction, which represents a suspended
computation. Some of the alternative notions of dynamic lifting that
appear in the literature make dynamic lifting part of the circuit
language, allowing circuits containing dynamic lifting to be
boxed. This turns circuits into thunks.

In recent work, {\cite{LeePVX21}} extended Proto-Quipper with a
version of dynamic lifting. They work with a single runtime modeled by
a category of \textit{quantum channels}, which are generalizations of
quantum circuits with a notion of branching for measurement results. A
quantum channel is a list of gates like a quantum circuit, with the
important exception that if the current gate is a measurement, the
list has two tails, one for each possible measurement outcome.
Consequently, the channels of Lee et al.\@ must either be implemented
as thunks, or as data structures that are exponentially large.  The
main difference with our work is that in our setting, dynamic lifting
ensures that boxed circuits are data structures that contain only one
branch (namely, the one corresponding to the actual measurement result
when the circuit is run), whereas in Lee et al.'s setting, either all
branches are evaluated, or the circuit is a thunk.

Another version of Proto-Quipper incorporating a form of dynamic lifting was
proposed by \cite{ColledanL22}. Their language uses a very general
version of dynamic lifting, which is even more general than the one present in
the Quipper language, and allows for measurements to be conditional
on the outcomes of prior measurements. As a consequence, the output
\emph{type} of their circuit can depend on the outcomes of the
measurements specified in the computation. They also work with
a single runtime where dynamic lifting is part of
their generalized notion of quantum circuits. While this alternative
notion of dynamic lifting is interesting in its own right, their
language does not come equipped with a denotational semantics.

QWire \cite{paykin2017qwire} is a quantum programming language that
also supports dynamic lifting. QWire has a host language and a circuit
language. The host language describes the computation of the classical
computer, while the circuit language describes the computation of the
quantum computer. QWire has a denotational semantics for the circuit
language, but not for the host language.  Dynamic lifting is part of
the syntax in the circuit language. Therefore QWire's notion of
quantum circuits differs from Proto-Quipper's notion. Besides dynamic
lifting, QWire also has a notion of \textit{static lifting} in the
form of a ``run'' function. This allows measuring all the qubits in a
circuit, returning boolean values to the host language, without
leaving any unmeasured quantum state. By contrast, Proto-Quipper-Dyn
does not require a run function, since it does not have separate host
and circuit languages. All circuits that are not constructed inside a
box are automatically executed, and dynamic lifting can be used to
bring measurement results into the control flow of the language.

We use enriched category theory to describe our categorical model for
dynamic lifting. There are some existing works that also use enriched
categories in the context of quantum programming languages. For
example, \cite{BMZ2018} use CPO-enrichment to model a version of
Proto-Quipper-M with recursion.  The main difference between our model
and that of Lindenhovius et al.\@ is that our model accounts for
dynamic lifting while their model accounts for recursion.
\cite{Rennela20} give a categorical model for a QWire-like language
that also uses enriched categories. Their language allows boxing a
circuit that uses dynamic lifting, which is quite different from how
boxing works in Proto-Quipper. As a result, circuits in their setting
are thunks and not data structures.  Also, in Rennela and Staton's
EWire language, the host language does not include wire types such as
a type of qubits, whereas Proto-Quipper does not have separate host
and circuit languages, and includes all types in a single
language. Consequently, Proto-Quipper has a linear type system,
whereas the EWire host language does not. This difference is also
reflected in the model: in Rennela and Staton's semantics, programs
are interpreted in a cartesian-closed category, whereas in our model,
they are interpreted in a monoidal category.

The fact that Proto-Quipper has two distinct runtimes (circuit
generation time and circuit execution time) suggests a possible
connection to another computational paradigm that also has multiple
runtimes, namely \emph{multi-staged
  computation}~\cite{TAHA2000211,MetaOCaml}. However, there are some
important differences.  One of them is that multi-staged computation,
such as in MetaML~\cite{TAHA2000211}, deals with potentially many
levels, but all of the levels share the same operations and the same
hardware; the primary purpose of staging is to precisely orchestrate
the order in which operations are evaluated. On the other hand, the
main purpose of dynamic lifting in Proto-Quipper is to interleave
computations from two different hardware models. The ``meta-language''
of Proto-Quipper terms has almost nothing in common with the ``object
language'' of circuits. Each of the two stages has its own distinct
operations: classical expressions and control flow for the
meta-language, and gates and measurements for the object language. In
particular, quantum circuits are not just code for expressions of the
meta-language.

Finally, we will mention the quantum programming language Silq
{\cite{silq}}. Like Proto-Quipper-Dyn, Silq also uses modalities to
keep track of the use of certain operations. For example, the modality
``mfree'' in Silq is used to indicate whether a computation uses
measurement. The difference is that Silq is not a circuit description
language, so it does not have a notion of boxed circuits or dynamic
lifting.

\subsection{Contributions}
In this paper, we describe the syntax and type system of an extension
of Proto-Quipper with dynamic lifting, called Proto-Quipper-Dyn. The
type system uses a system of modalities to keep track of the use of
dynamic lifting. We also provide an operational semantics, using two
different kinds of configurations to model circuit generation time and
circuit execution time. We further provide an abstract categorical
semantics for this language, in which dynamic lifting is modeled by a
map $\Bit \to T\Bool$, where $T$ is a monad encapsulating circuit
execution. By an ``abstract'' categorical semantics, we mean that we
only state the properties that a categorical model must satisfy to
give a sound interpretation of the language, without constructing an
actual concrete example of such a model. We give such a concrete model
in a companion paper \cite{FKRS-model-2022}.

The rest of the paper is organized as follows: In Section
\ref{sec:model}, we briefly recall the basics of enriched category
theory, and then we give an axiomatization of a general categorical
semantics for dynamic lifting.  In Section \ref{sec:types}, we define
a type system for dynamic lifting that uses a system of modalities.
We then show how a typing judgment with modalities is interpreted as a
morphism in our categorical model.  In Section \ref{sec:op}, we define
a call-by-value big-step operational semantics for our language.  We
show that the operational semantics satisfies type preservation and
that the type system guarantees error freeness. We also show that the
operational semantics is sound with respect to the enriched
categorical semantics.  In Section \ref{sec:app}, we give some
applications of dynamic lifting in Proto-Quipper-Dyn. We finish the paper
with some concluding remarks in Section \ref{sec:conclude}.

\section{An enriched categorical semantics for dynamic lifting}
\label{sec:model}
In this section we will give a general categorical semantics for
dynamic lifting. Our categorical semantics is based on enriched categories,
which are generalizations of ordinary categories.
In enriched categories, instead of hom-sets, one works with hom-objects, which are objects in
a monoidal category. 
\begin{definition}
  Let $\V$ be a monoidal category. 
  A \emph{$\V$-enriched category} $\A$ (or \emph{$\V$-category} for short) is given by the following:
  \begin{itemize}
  \item A class of objects, also denoted $\A$. 
  \item For any $A, B \in \A$, an object $\A(A, B)$ in $\V$.
  \item For any $A \in \A$, a morphism in $u_{A} : I \to \A(A, A)$ in
    $\V$, called the \emph{identity} on $A$.
  \item For any $A, B, C \in \A$, a morphism $c_{A,B,C} : \A(A, B)
    \otimes \A(B, C) \to \A(A, C)$ in $\V$, called
    \emph{composition}.
  \item The composition and identity morphisms must satisfy suitable diagrams in $\V$ (see \cite{kelly1982basic,borceux1994handbook2}).
  \end{itemize}
\end{definition}

\begin{remarks}
\begin{itemize}
\item Many concepts from non-enriched category theory can be generalized to the enriched setting. For example,
  $\V$-functors, $\V$-natural transformations, $\V$-adjunctions and
  the $\V$-Yoneda embedding are all
  straightforward generalizations of their non-enriched counterparts. We refer to \cite{kelly1982basic,borceux1994handbook2}
  for comprehensive introductions. 
\item In the rest of this paper, when we speak of a map $f : A \to B$ in a $\V$-enriched category $\A$, we mean a morphism of the form $f : I \to \A(A, B)$ in $\V$. Furthermore, when $g : B\to C$ is another map in $\A$, we write $g \circ f : A \to C$ as a shorthand for
  \[ I \xrightarrow{f\otimes g} \A(A, B)\otimes \A(B, C) \xstackrel{c}{\to} \A(A, C).\]

\item A $\V$-enriched category $\A$ gives rise to an ordinary category $V(\A)$, called the \textit{underlying category}\footnote{$V$ stands for ``underlying'' because the letter $U$ serves another purpose in this paper.} of $\A$, where the objects of $V(\A)$
  are objects of $\A$ and a hom-set is defined as $V(\A)(A, B) := \V(I, \A(A, B))$ for any $A, B\in V(\A)$. Similarly,
  a $\V$-functor $F : \A \to \mathbf{B}$ gives rise to an ordinary functor $VF : V(\A) \to V(\mathbf{B})$ and a $\V$-natural transformation $\alpha : F \to G$
  gives rise to an ordinary natural transformation $V\alpha : VF \to VG$. 
  
\end{itemize}
\end{remarks}

Ordinary symmetric monoidal categories can be generalized to
 enriched categories as well. 
\begin{definition}
  Let $\V$ be a symmetric monoidal category. A $\V$-category $\A$ is
  symmetric monoidal if it is equipped with the following:
  \begin{itemize}
  \item There is an object $I \in \A$ called the \emph{tensor unit}.
    For any $A, B\in \A$, there is an object $A \otimes B\in \A$. Moreover,
    for any $A_{1}, A_{2}, B_{1}, B_{2} \in \A$, there is a morphism
    \[\mathrm{Tensor} : \A(A_{1}, B_{1})\otimes \A(A_{2}, B_{2}) \to \A(A_{1}\otimes A_{2}, B_{1}\otimes B_{2})\]
in $\V$.
    The tensor product is a bifunctor in the sense that $\mathrm{Tensor} \circ (u_{A}\otimes u_{B}) = u_{A\otimes B}$ for the identity maps $u_{A}, u_{B}, u_{A\otimes B}$, and the following diagram commutes for any $A_{1}, A_{2}, B_{1}, B_{2}, C_{1}, C_{2}\in \A$. 
    \[
      \begin{tikzcd}
        \A(A_{1}, B_{1})\otimes \A(A_{2}, B_{2}) \otimes \A(B_{1}, C_{1})\otimes \A(B_{2}, C_{2})
        \arrow[r, "c\otimes c"]
        \arrow[d, "\mathrm{Tensor}\otimes \mathrm{Tensor}"]& \A(A_{1}, C_{1})\otimes \A(A_{2}, C_{2}) \arrow[d, "\mathrm{Tensor}"]\\
        \A(A_{1}\otimes A_{2}, B_{1}\otimes B_{2}) \otimes \A(B_{1}\otimes B_{2}, C_{1}\otimes C_{2})\arrow[r, "c"]& \A(A_{1}\otimes A_{2}, C_{1}\otimes C_{2})
      \end{tikzcd}
    \]
  \item There are the following $\V$-natural isomorphisms in $\A$ and they satisfy the same coherence diagrams for symmetric
    monoidal categories. 
    \[l_{A} : I\otimes A \to A\]
    \[r_{A} : A\otimes I \to A\]
    \[\gamma_{A, B} : A\otimes B \to B\otimes A\]
    \[\alpha_{A, B,C} : (A\otimes B)\otimes C \to  A \otimes (B\otimes C)\]
  \end{itemize}
\end{definition}

If the $\V$-category $\A$ is symmetric monoidal, then its underlying category $V(\A)$ is symmetric monoidal. 
For any maps $f : A_{1} \to B_{1}, g : A_{2} \to B_{2}$ in $\A$, we write the map $f \otimes g : A_{1}\otimes A_{2} \to B_{1}\otimes B_{2}$ as a shorthand for the
following composition.
\[ I \xstackrel{f\otimes g}{\to} \A(A_{1}, B_{1})\otimes \A(A_{2}, B_{2}) \xstackrel{\mathrm{Tensor}}{\to} \A(A_{1}\otimes A_{2}, B_{1}\otimes B_{2})\]

\subsection{An axiomatization of enriched categorical models of dynamic lifting}
In the following, we assume $\V$ to be a cartesian closed category with coproducts.
For any $A, B \in \V$, we write $A \times B$ for the
cartesian product, $A \Rightarrow B$ for the exponential object, and $1 \in \V$ for the terminal object.
Since $\V$ is cartesian closed, it is self-enriched, i.e., $\V$ is a
$\V$-category where the hom-objects are
defined by $\V(A, B) := A \Rightarrow B$.

We will now focus on defining a $\V$-enriched category $\A$ that
models dynamic lifting.  We give a sequence of definitions that
specify a sequence of properties (\ref{closed})-(\ref{dynlift}), which
will culminate in Definition~\ref{def:abstract} of a model for
Proto-Quipper with dynamic lifting.

\begin{definition}
  \label{def:linear-nonlinear}
  A $\V$-category $\A$ is a \textit{linear-non-linear programming language model} if
  \begin{enumerate}
    \setcounter{enumi}{\value{equation}}
  \item \label{closed} $\A$ has coproducts and is symmetric monoidal closed,
    i.e., it is symmetric monoidal and there is a $\V$-adjunction $- \otimes A \dashv A \multimap -$
    for each $A \in \A$.
  \item \label{lin-nonlin} $\A$ is equipped with a $\V$-adjunction
    \[p : \V \to \A \dashv \flat : \A \to \V\]
    such that $p$ is a strong monoidal $\V$-functor.
    \setcounter{equation}{\value{enumi}}
  \end{enumerate}
\end{definition}

\begin{remarks}
\begin{itemize}
\item The requirement that $\A$ has coproducts and is symmetric monoidal closed implies that
  it can model function types and sum types in a functional programming language. Moreover, since $- \otimes A$ is a left adjoint $\V$-functor for any $A \in \A$, it preserves the coproducts, so the tensor products distribute over coproducts in $\A$. 
  
\item The adjunction in (\ref{lin-nonlin}) is often called a
  \textit{linear-non-linear adjunction} \cite{benton1994mixed}. Here,
  the assumption that $p$ is a strong monoidal $\V$-functor means that
  there exist isomorphisms $e : I \to p1$ and $m : pX \otimes pY \to
  p(X \times Y)$ making some diagrams commute (see Appendix
  \ref{app:strong-monoidal}).
\item Since $p$ is strong monoidal and $\V$ is cartesian, for any $X
  \in \V$, there are maps $\mathsf{discard}_{X}: pX \to I$ and
  $\mathsf{dup}_{X} : pX\to pX\otimes pX$ in $\A$.  Moreover, for any
  map $f : X \to Y$ in $\V$, we have the following in $\A$.
  \[ \mathsf{dup}_{Y}\circ pf = (pf \otimes pf) \circ \mathsf{dup}_{X}\]
  We call objects of the form $pX \in \A$ \textit{parameter objects},
  since they can be duplicated and discarded. For example, $\Bool := I
  + I \cong p1 + p1 \cong p(1+1)$ is a parameter object.
  
\item For any $X \in \V, B\in \A$, we write $\delta$ for the
  isomorphism $\delta: \A(pX, B) \cong \V(X, \flat B)$, and
  $\force_{B}$ for the counit $\force_{B} : p\flat B \to B$.
\end{itemize}
\end{remarks}

\begin{definition}\label{def:convex-space}
  A \emph{convex space} is a set $X$ equipped with a \emph{convex sum}
  operation, which assigns to any $x,y\in X$ and $p,q\in[0,1]$ such
  that $p+q=1$ an element $px+qy\in X$, subject to certain standard
  conditions, which are detailed in Appendix \ref{app:def-convex}.  A
  category is \emph{enriched in convex spaces} if each hom-set is
  equipped with the structure of a convex space, and moreover,
  composition is \textit{bilinear} with respect to convex sum, i.e.,
  $(pf + qg) \circ h = p(f\circ h) +
  q(g\circ h)$ and $h \circ (pf + qg) =
  p(h \circ f) + q(h \circ g)$.
\end{definition}

As mentioned in the introduction, Proto-Quipper-Dyn is parameterized
by two (ordinary) small categories $\m$ and $\q$ of \emph{circuits} and
\emph{quantum operations}, respectively. We now
specify the properties that these categories must satisfy.
\begin{assumption}
  \label{ass:mq}
  We assume that we are given two small symmetric monoidal categories $\m$
  and $\q$, satisfying the following properties:
  \begin{itemize}
  \item[(1)] $\m$ and $\q$ have the same objects, including a distinguished object called $\Bit$. 
    The category $\m$ has distinguished morphisms $\mathrm{zero}, \mathrm{one} : I \to \Bit$. 
    
  \item[(2)] $\q$ has a coproduct $\Bit = I + I$, and the tensor product in $\q$ distributes over this coproduct.

  \item[(3)] There exists a given strict symmetric monoidal functor $J : \mathbf{M} \to \q$ that is the identity on objects and $J(\mathrm{zero}) = \mathrm{inj}_{1} : I \to I + I, J(\mathrm{one}) = \mathrm{inj}_{2} : I \to I + I$. We call $J$ the \textit{interpretation functor}.
    
  \item[(4)] The category $\q$ is enriched in convex spaces.
    
  \item[(5)] For any $A\in \q$, and $f : I \to \Bit\otimes A \in \q$, we have 
    $f = p_{1} (\mathrm{inj}_{1} \otimes f_{1} ) + p_{2}(\mathrm{inj}_{2} \otimes f_{2} )$,
    where $\mathrm{inj}_{1}, \mathrm{inj}_{2} : I \to I + I$ and $p_1, p_2 \in [0, 1]$
    are uniquely determined real numbers such that $ p_1 + p_2 = 1$. When $p_{i} \not = 0$, the map $f_{i} : I \to A$ is also unique. 
  \end{itemize}
\end{assumption}

The categories $\m$ and $\q$ are not only used in the categorical semantics,
but also in the operational semantics of Proto-Quipper-Dyn (i.e., to run
the program, we must know what a circuit is and what a quantum
operation is). Therefore, these categories should be regarded as given
as part of the language specification, rather than as a degree of
freedom in the semantics. On the other hand, nothing in the
operational or denotational semantics depends on particular properties
of $\m$ and $\q$ other than properties (1)--(5) above. Therefore,
Proto-Quipper-Dyn can handle a wide variety of possible circuit models and
physical execution models.

In practice, the category $\m$ will be a category of quantum circuits
and the category $\q$ will be a category of quantum operations. These
categories will typically have additional objects, such as $\Qubit$
and perhaps $\Qutrit$, and additional morphisms, such as $H : \Qubit
\to \Qubit$ and $\Meas : \Qubit \to \Bit$.
Assumption (5) means that any morphism with domain $I$ and a bit state in its codomain
is a convex sum of two morphisms. This property is used in the rule
for dynamic lifting in the operational semantics.

  \begin{definition}
    Suppose the $\V$-enriched category $\A$ is a linear-non-linear programming model. We say it
    supports \textit{box-unbox operations} if the following hold.
    \begin{enumerate}
    \setcounter{enumi}{\value{equation}}
   
  \item \label{circ-embedding} There is a fully faithful embedding
    $\psi : \mathbf{M} \stackrel{\psi}{\hookrightarrow} V(\A)$ and $\psi$ is strong monoidal.
    \item \label{box-unbox} Let $\mathcal{S}$ denote the set of
      objects in the image of $\psi$.
    For any $S, U\in \mathcal{S}$, there is an isomorphism
    \[ \flat(S \multimap U) \stackrel{e}{\cong}\A(S, U).\]
    \setcounter{equation}{\value{enumi}}
    \end{enumerate}
  \end{definition}

  Condition (\ref{circ-embedding}) implies that there is a circuit subcategory in $\A$.
  Using condition (\ref{box-unbox}), we define $\boxt = p(e)$ and $\unboxt = p(e^{-1})$,
  and there is an isomorphism $p\flat(S \multimap U) \stackrel{\boxt/ \unboxt}{\cong} p\A(S, U)$. 
  Elements of $p\A(S, U)$ correspond to boxed circuits with input $S$ and output $U$. 

  If a $\V$-enriched category $\A$ satisfies (\ref{closed})--(\ref{box-unbox}), then it is a model
  for Proto-Quipper \textit{without} dynamic lifting. For example, the $\set$-enriched category $\mbbar$ in \cite{RS2017-pqmodel}
  is such a model.
  To support dynamic lifting, we define the following monad to account for the category $\q$.   

  \begin{definition}
    Let $\A$ be a symmetric monoidal $\V$-category and let $T : \A \to \A$ be a $\V$-monad on $\A$.
    We say $T$ is a \textit{dynamic lifting monad} if the following hold.
    \begin{enumerate}
      \setcounter{enumi}{\value{equation}}
      
    \item \label{monad-t} $T$ is a commutative strong $\V$-monad. 
      For any $A, B\in \A$, we write $t_{A,B} : A\otimes TB \to T(A\otimes B)$ for the strength 
      and $s_{A,B} : TA \otimes B \to T(A\otimes B)$ for the costrength.
    \item \label{convexity}
      Let $V(\A)$ be the underlying category of $\A$,
      let $VT$ be the underlying monad of $T$,
      and let $\Kl_{VT}(V(\A))$ be the Kleisli category of $VT$.  The Kleisli category $\Kl_{VT}(V(\A))$ is
      enriched in convex spaces. 

    \item \label{embeddings}

      There are the following
      fully faithful embeddings:
      \[\mathbf{M} \stackrel{\psi}{\hookrightarrow} V(\A),\]
      \[\mathbf{Q} \stackrel{\phi}{\hookrightarrow} \Kl_{VT}(V(\A)).\]
      These embedding functors are strong monoidal, and $\phi$ preserves
      the convex sum. 
      Moreover, the following
      diagram commutes for any $S, U \in \m$. 

      \[
        \begin{tikzcd}
          \mathbf{M}(S, U)
          \arrow[r, "\psi_{S, U}"]
          \arrow[d, "J_{S, U}"]
          & V(\A)(S, U) \arrow[d, "E_{S, U}"]\\
          \mathbf{Q}(S, U) \arrow[r, "\phi_{S, U}"]& \Kl_{VT}(V(\A))(S, U)
        \end{tikzcd}
      \]
      Here $E : V(\A) \to \Kl_{VT}(V(\A))$ is the
      the functor such that
      $E(A) = A$ and $E(f) = \eta \circ f$.

    \item \label{dynlift}
      There are maps $\Dyn : \Bit \to T\Bool$ and $\mathrm{init} : \Bool \to \Bit$ in $\A$
      such that the following diagram commutes.
      \[
        \begin{tikzcd}
          & \Bit \arrow[d, "\Dyn"]\\
          \Bool \arrow[r, "\eta"] \arrow[ur, "\mathrm{init}"]  & T\Bool
        \end{tikzcd}
      \]
      \setcounter{equation}{\value{enumi}}
    \end{enumerate}
  \end{definition}

\begin{remarks}
\begin{itemize}
\item 
  The objects of the Kleisli category $\Kl_{VT}(V(\A))$ are the same
  as the objects of $\A$, and
  the hom-set is given by $\Kl_{VT}(V(\A))(A, B) := V(\A)(A, VTB) = \V(1, \A(A, TB))$ for any $A, B \in \A$.
  Moreover, $\V(1, \A(A, TB)) = \V(1, \Kl_{T}(\A)(A, B)) = V(\Kl_{T}(\A))(A, B)$. 

\item Note that in condition (\ref{convexity}), we are not taking a
  $\V$-enriched Kleisli category of the $\V$-monad $T$, but just an
  ordinary Kleisli category of the ordinary monad $VT$. Thus, the
  Kleisli category is not $\V$-enriched. However, we do require it to be
  enriched in convex spaces, which amounts to requiring the existence
  of additional operations on its hom-sets, in the sense of Definition~\ref{def:convex-space}.
  \item Since $T$ is a commutative strong $\V$-monad, $VT$ is a
    commutative strong (ordinary) monad. Therefore the Kleisli category $\Kl_{VT}(V(\A))$ is monoidal.
    For any $f : A_{1} \to VTB_{1}$ and $g : A_{2} \to VTB_{2}$ in $\Kl_{VT}(V(\A))$,
    we define $f\otimes g \in \Kl_{VT}(V(\A))(A_{1}\otimes A_{2}, B_{1}\otimes B_{2})$ to be the following 
    \[ A_{1}\otimes A_{2} \xstackrel{f\otimes g}{\to} VTB_{1}\otimes VTB_{2}
      \xstackrel{s}{\to} VT(B_{1}\otimes VTB_{2})
      \xstackrel{Tt}{\to} VTVT(B_{1}\otimes B_{2})
      \xstackrel{\mu}{\to} VT(B_{1}\otimes B_{2}).
    \]

  \item
    Condition (\ref{embeddings}) expresses the requirement that the enriched category $\A$
  must combine
  both categories $\m$ and $\q$, i.e., they are subcategories
  of $V(\A)$ and its Kleisli category, respectively. Thus $\A$ has both quantum circuits and quantum operations. The commutative diagram implies that
  a circuit in $\A$ can be used as a quantum operation.
    
  \item Since $\psi(S) = \phi(S)$ for any $S \in \m, \q$, 
    we define $\Bit = \psi(\Bit) = \phi(\Bit) \in \A$.
    
  \item Condition (\ref{dynlift}) gives a categorical characterization of dynamic lifting.  
    The map $\Dyn$ is not in the image of $\phi$ or $\psi$, and therefore it is neither
    a quantum circuit nor a quantum operation. 
\end{itemize}
\end{remarks}

\begin{definition}
  \label{def:abstract}
  We say a $\V$-enriched category $\A$ is a \emph{model for Proto-Quipper with dynamic lifting} if it satisfies
  (\ref{closed})--(\ref{dynlift}). 
\end{definition}

We have now axiomatized a general categorical model for Proto-Quipper with dynamic lifting. In \cite{FKRS-model-2022},
we give a construction of a concrete model based on \textit{biset-enrichment} that satisfies (\ref{closed})-(\ref{dynlift}). 
In the rest of this paper, we will be focusing on
showing this abstract categorical model $\A$ is sound with respect to
the type system and the operational semantics.

\section{A type system for dynamic lifting}
\label{sec:types}

In this section, we present the syntax of Proto-Quipper-Dyn and a type system for dynamic lifting. 
Our typing judgments have the form $\Gamma \vdash_{\alpha} M : A$, where $\alpha ::= 0 \ | \ 1$ is a modality used to keep track of dynamic lifting.
When $\alpha = 1$, the term $M$ is guaranteed not to perform any
dynamic lifting operations while it is being reduced to a value. Such
computations can therefore be carried out at circuit generation time. When $\alpha = 0$, $M$ may invoke dynamic lifting so the evaluation of $M$ needs to be performed at circuit execution time.

\begin{definition}[Syntax]
The syntax of Proto-Quipper-Dyn is in Figure \ref{fig:syntax}. 
  \begin{figure}
    \centering
    \[ \small\def\arraystretch{1.1}
    \begin{array}{llll}
      \\
      \textit{Modality} & \alpha, \beta &::=& 0 \mid 1
      \\ 
      \textit{Types} & A, B  &::=&  \mathbf{Unit} \mid  \mathbf{Qubit} \mid
      \Bit \mid  \Bool\mid  {!}_\alpha A 
      \mid A \multimap_\alpha B \mid  \mathbf{Circ}(S, U) \mid  A \otimes B
      \\ 

      \textit{Parameter Types} & P, R  &::=& \mathbf{Unit} \mid  \mathbf{Nat} 
      \mid {!}_\alpha A \mid  \mathbf{Circ}(S, U) \mid P \otimes R

      \\ 
      
      \textit{Simple Types} & S, U  &::=& \mathbf{Unit} \mid  \mathbf{Qubit}\mid  \Bit 
      \mid S \otimes U
      
      \\ 
      
      \textit{Terms} & M, N  &::=& c\mid  x \mid  \lambda x . M \mid  M \ N \mid  \mathsf{Unit} \mid  (a, \cC, b) \mid  \mathsf{apply}(M, N) \mid  \force M \\

      &&& {~}\mid \mathsf{lift}\ M \mid  \mathsf{box} \ U \ M \mid  (M, N) \mid  \mathsf{let}\ (x, y) = N\  \mathsf{in}\ M \mid  \Dyn M

      \\ 
      
      \textit{Simple Terms} & a, b  &::=& \ell  \mid  \mathsf{Unit}\mid  (a, b)

      \\ 
      \textit{Contexts} & \Gamma  &::=& \cdot \mid  x : A, \Gamma \mid  \ell : \mathbf{Qubit}, \Gamma \mid \ell : \Bit, \Gamma

      \\ 
      \textit{Parameter contexts} & \Phi  &::=& \cdot \mid  x : P, \Phi.
      
      \\ 
      \textit{Label Contexts} & \Sigma  &::=& \cdot \mid  \ell : \mathbf{Qubit}, \Sigma \mid  \ell : \Bit, \Sigma

      \\ 

      \textit{Values}
      & V &::=& x \mid  \ell \mid  \lambda x. M \mid  \mathsf{lift}\ M 
      \mid (a, \cC, b) \mid  (V, V') \mid  \mathsf{Unit}

      \\ 
      
      \textit{Circuits} & \multicolumn{3}{l}{\cC, \dD : \Sigma \to \Sigma'}
\end{array}
    \]
    \caption{The syntax for Proto-Quipper-Dyn}
    \label{fig:syntax}
  \end{figure}

\end{definition}

The modality $\alpha$ appears in the linear function type $A \multimap_{\alpha} B$ and
the linear exponential type ${!}_{\alpha} A$. This is because the values of $A \multimap_{\alpha} B$ and ${!}_{\alpha} A$
are \textit{thunks} and we use the modality $\alpha$ in the types to keep track of
the dynamic lifting within the thunks. $\mathbf{Circ}(S, U)$ denotes
a type of circuits with input $S$ and output $U$. The values of this
type are boxed quantum circuits. They can be further manipulated by
meta-operations such as circuit reversal, circuit iteration, or
printing; these operations are treated as constants in the
language, i.e., we do not fix a particular set of such operations, but
assume that they would be defined in a standard library that comes
with any particular instance of Proto-Quipper-Dyn.

The terms of our language are similar to the ones from 
\cite{RS2017-pqmodel}, with the addition of a term construct for dynamic lifting $\Dyn M$, which will
be evaluated to a boolean value. The term $c$ ranges over constants
such as booleans, natural numbers, and built-in functions. A term of parameter type can be duplicated or discarded.
A value of simple type corresponds to a state. Our language and semantics can accommodate coproducts (sum types), but we elide the treatment here
for the sake of simplicity. 

We make a distinction between variables and labels.
 A label $\ell$ corresponds
 to a wire in a circuit, or to an
 address of a bit or qubit state. Consequently, a label is a value that can only have type $\Bit$ or $\Qubit$. Labels can only be renamed, not substituted.  
 Every label context $\Sigma$ has an obvious interpretation $\interp{\Sigma}$ in the category $\m$ as a tensor of the appropriate sequence of the objects $\Qubit$ and $\Bit$.
 We write $\dD : \Sigma \to \Sigma'$ to denote a quantum circuit, i.e., a morphism $\dD : \interp{\Sigma} \to \interp{\Sigma'}$.

\begin{definition}[Typing] 
  \label{typing}
  The typing rules are in Figure \ref{fig:typing}. 

  \begin{figure}
    \centering
    \[ \small
    \begin{tabular}{llll}
      \\
      \infer[\textit{var}]
      {\Phi, x : A \vdash_1 x : A}{}
      &
        \infer[\textit{label}]
        {\ell : \mathbf{Qubit}|\Bit \vdash_1 \ell : \mathbf{Qubit}|\Bit}{}
      \\
      \\
      \infer[\textit{app}]
      {\Gamma_1 + \Gamma_2 \vdash_{\alpha_1 \& \alpha_2 \& \beta} M N : B}
      {\Gamma_1 \vdash_{\alpha_1} M :  A \multimap_\beta B & \Gamma_2 \vdash_{\alpha_2} N : A}

                                                           &
                                                             \infer[\textit{lambda}]{\Gamma \vdash_{1} \lambda x . M : A \multimap_\alpha B}
                                                             {\Gamma, x : A \vdash_{\alpha} M : B} %
      \\
      \\
      \infer[\textit{lift}]{\Phi \vdash_1 \mathsf{lift}\ M : {!}_\alpha A}
      {\Phi \vdash_\alpha M : A}
      &
        \infer[\textit{force}]{\Gamma \vdash_{\alpha \& \beta} \mathsf{force}\ M : A}
        {\Gamma \vdash_\beta M :\ !_\alpha A}
      \\
      \\

      \infer[\textit{box}]{\Gamma \vdash_\alpha \mathsf{box}\ S \  M : \mathbf{Circ}(S,U)}{\Gamma \vdash_\alpha M : {!}_1 (S \multimap_1 U)}
      &
        \infer[\textit{apply}]{\Gamma_1 + \Gamma_2 \vdash_{\alpha \& \beta} \mathsf{apply}(M, N) :  U}{\Gamma_1 \vdash_\alpha M : \mathbf{Circ}(S,U) & \Gamma_2 \vdash_\beta N : S}
      \\
      \\
      \infer[\textit{circ}]{\Phi \vdash_1 (a, \cC, b) :  \mathbf{Circ}(S, U)}
      {
      \begin{array}{ll}
        \Sigma_1 \vdash_1 a : S
        &
          \Sigma_2 \vdash_1 b : U \\
        \cC : \Sigma_1 \to \Sigma_2
      \end{array}
      }
      &
        \infer[\textit{dynlift}]{\Gamma \vdash_0 \Dyn\ M  : \Bool}
        {\Gamma \vdash_\alpha M  : \Bit}
      \\
      \\
      \infer[\textit{pair}]
      {\Gamma_1 + \Gamma_2 \vdash_{\alpha_1 \& \alpha_2} (M, N) : A\otimes B}
      {\Gamma_1 \vdash_{\alpha_1} M :  A & \Gamma_2 \vdash_{\alpha_2} N : B}

                                                           &
                                                             \infer[\textit{let}]
                                                             {\Gamma_1 + \Gamma_2 \vdash_{\alpha_1 \& \alpha_2} \mathsf{let}\ (x, y) = N\  \mathsf{in}\ M : C}
                                                             {\Gamma_1, x : A, y : B \vdash_{\alpha_1} M :  C &  \Gamma_2 \vdash_{\alpha_2} N : A\otimes B}
                                                                                                                
    \end{tabular}
    \]
    \caption{The typing rules for Proto-Quipper-Dyn}
    \label{fig:typing}
  \end{figure}
  \end{definition}

We write $\alpha \& \beta$ for the boolean conjunction of $\alpha$ and $\beta$ so that, e.g., $0 \& 1 = 0$. If $\Gamma_{1} = \Phi, \Gamma_{1}'$ and $\Gamma_{2} = \Phi, \Gamma_{2}'$, we write $\Gamma_{1} + \Gamma_{2}$ for $\Phi, \Gamma_{1}', \Gamma_{2}'$. 

In the \textit{var} rule, we require a parameter context $\Phi$.  
In the \textit{lift} and \textit{lambda} rules, the modality $\alpha$ is moved to the type and the current modality (i.e., modality in the conclusion) is set
to $1$. This is because the lift and lambda terms are values, and values do not perform dynamic lifting.
In fact, all values have modality $1$. 

In elimination rules such as \textit{app} and \textit{force}, the modality in the type affects the current modality of the typing judgment through boolean conjunction. This is related to how the evaluations are
performed for these terms. For example, when evaluating
the term $M N$, we will first evaluate $M$, then evaluate $N$ and
finally perform a beta-reduction. Thus, the evaluation of $MN$ could
perform dynamic lifting of $\alpha_{1}=0$, $\alpha_{2}=0$, or
$\beta=0$. Consequently,
 the modality for the typing judgment of $M N$ is the boolean conjunction of all these related modalities. 

By the \textit{dynlift} rule, an application of dynamic lifting sets the current modality to $0$, signifying that a dynamic lifting is performed.
In the \textit{box} rule, a term $M$ can only be
boxed into a circuit if it has type $!_{1}(S \multimap_{1} U)$. This ensures that the value of $M$ (denoted by $V$) does not use dynamic lifting.
Thus, when evaluating the term $(\boxt S\ V)$, a dynamic lifting cannot occur. This prevents a class of runtime errors in Quipper that are caused by boxing functions that use dynamic lifting. 

In the \textit{apply} rule, depending on the modality $\alpha_{1}\& \alpha_{2}$, the term $\mathsf{apply}(M, N)$ either appends the
quantum circuit $M$ to $N$, which is done at circuit generation time, or applies the quantum operation $M$ to $N$, which is done at circuit execution time. 
The \textit{circ} rule defines a well-typed quantum
circuit. In practice, we often assume that a set of well-typed quantum
gates is provided as pre-defined constants of the language, so that the programmer does not need to use the \textit{circ} rule. 

The following lemma shows that a value can only have modality $1$ and, in particular, that the free variables of a parameter must come from a parameter context. 
\begin{lemma}
  \label{lem:value}
  If $\Gamma \vdash_\alpha V : B$, then $\alpha = 1$. Moreover, if $\Gamma \vdash_\alpha V : P$, then
  $\alpha = 1$ and $\Gamma = \Phi$.
\end{lemma}

The following lemma shows that the type system has the usual substitution property. 
\begin{lemma}[Substitution]
  If $\Gamma_1, x : A, \Gamma_1' \vdash_\alpha M : B$ and $\Gamma_2 \vdash_{1} V : A$, then $\Gamma_1, \Gamma_1', \Gamma_2 \vdash_{\alpha} [V/x]M : B$.
\end{lemma}

\subsection{Interpretation of the typing rules}
The modality is a syntactic device to track the dynamic lifting monad $T$. 
 We will interpret $\Gamma \vdash_{1} M : A$ as a map $\interp{\Gamma}\to \interp{A}$ in $\A$, and $\Gamma \vdash_{0} M : A$ as a map $\interp{\Gamma}\to T\interp{A}$.
 The modalities in types such as $A \multimap_\alpha B$ and
 ${!}_\alpha A$  also indicate occurrences of the
 dynamic lifting monad $T$.
\begin{definition}
  We interpret types as objects in $\A$.
  \[\def\arraystretch{1.1}
    \begin{array}{lll}
      \interp{A \multimap_1 B} &=& \interp{A} \multimap \interp{B}\\
      \interp{A \multimap_0 B} &=& \interp{A} \multimap T \interp{B}\\
      \interp{A\otimes B} &=&  \interp{A}\otimes \interp{B}\\
      \interp{!_1 A} &=& p\flat\interp{A}\\
      \interp{!_0 A} &=& p\flat T \interp{A}\\
      \interp{\mathbf{Circ}(S, U)} &=& p\A(\interp{S}, \interp{U})\\
      \interp{\Bool} &=& p(1+1)\\
      \interp{\Bit} &=& \Bit\\
      \interp{\Qubit} &=& \Qubit\\
    \end{array}
  \]
\end{definition}

For a parameter type $P$, there exists $X \in \V$ such that $\interp{P} = pX$. 
For a simple type $S$, there exists $Y\in \m$ such that $\interp{S} = \psi Y$. We call objects of the form $\psi Y$ \textit{simple objects}. We write $\alpha \interp{A}$ to mean $T \interp{A}$ if $\alpha = 0$, otherwise it is $\interp{A}$.
We interpret a context $\Gamma$ as a tensor product of all objects in $\Gamma$ (denoted by $\interp{\Gamma}$). The interpretation of parameter context $\interp{\Phi}$ is a parameter object and the interpretation of a label context $\interp{\Sigma}$ is a simple object. 
Without loss of generality, we assume that if $\interp{\Sigma}= \interp{\Sigma'}$, then $\Sigma = \Sigma'$ (this condition can always be ensured by making additional isomorphic copies of objects, if necessary).

The interpretation of typing judgements is defined as follows.
\begin{definition}[Interpretation]
  \label{def:interpretation}
  To each valid typing judgement $\Gamma \vdash_\alpha M : A$, we associate a map $\interp{M} : \interp{\Gamma} \to  \alpha \interp{A}$ in $\A$, called its \emph{interpretation}. Note that $\interp{M}$ here is an abbreviation for $\interp{\Gamma \vdash_\alpha M : A}$. 

  The interpretation is defined by induction on the derivation of
  $\Gamma \vdash_\alpha M : A $.  Here we show a few cases, the rest
  are in Appendix \ref{app:interpretation}.

  \begin{itemize}
    \item Case
      \begin{center}
        \begin{tabular}{l}
          \infer{\Gamma \vdash_0 \Dyn M  : \Bool.}
          {\Gamma \vdash_\alpha M  : \Bit}
        \end{tabular}
      \end{center}
      By induction hypothesis, we have $\interp{M} : \interp{\Gamma}\to \alpha \interp{\Bit}$.
      If $\alpha = 1$,
      we define $\interp{\Dyn M}$ by
      \[\interp{\Gamma} \xstackrel{\interp{M}}{\to} \interp{\Bit} \xstackrel{\Dyn}{\to} T\interp{\Bool}.\]

      If $\alpha = 0$,
      we define $\interp{\Dyn M}$ by
      \[\interp{\Gamma} \xstackrel{\interp{M}}{\to} T\interp{\Bit} \xstackrel{T \Dyn}{\to} TT\interp{\Bool} \xstackrel{\mu}{\to} T\interp{\Bool}.\]

    \item Case
      \begin{center}
        \begin{tabular}{l}
          \infer[]{\Gamma \vdash_{1} \lambda x . M : A \multimap_\alpha B.}
          {\Gamma, x : A \vdash_{\alpha} M : B}
        \end{tabular}
      \end{center}
      By induction hypothesis, we have $\interp{M} :\interp{\Gamma}\otimes \interp{A} \to \alpha\interp{B}$.
      Using monoidal closedness, we define $\interp{\lambda x.M} := \mathsf{curry}(\interp{M}) : \interp{\Gamma}\to  \interp{A} \multimap \alpha\interp{B}$.

    \item Case
      \begin{center}
        \begin{tabular}{l}
          \infer[\textit{app}]
          {\Phi, \Gamma_1, \Gamma_2 \vdash_{\alpha_1 \& \alpha_2 \& \beta} M N : B.}
          {\Phi, \Gamma_1 \vdash_{\alpha_1} M :  A \multimap_\beta B & \Phi, \Gamma_2 \vdash_{\alpha_2} N : A}
        \end{tabular}
      \end{center}
      Here we only consider the case where $\alpha_{1} = \alpha_{2} = \beta = 0$. The other cases are similar. 
      By induction hypothesis, we have morphisms $\interp{M} : \interp{\Phi}\otimes \interp{\Gamma_1} \to T (\interp{A} \multimap T \interp{B}))$ and $\interp{N} : \interp{\Phi}\otimes\interp{\Gamma_2} \to T \interp{A}$.
      Thus we define $\interp{MN}$ to be the following.

      \[\def\arraystretch{1.3}
        \begin{array}{l@{~}l}
        \interp{\Phi}\otimes \interp{\Gamma_1}\otimes \interp{\Gamma_2} &\xstackrel{\mathsf{dup}\otimes \interp{\Gamma_1}\otimes \interp{\Gamma_2}}{\to}\interp{\Phi}\otimes \interp{\Phi}\otimes \interp{\Gamma_1}\otimes \interp{\Gamma_2}
        \xstackrel{\interp{M}\otimes \interp{N}}{\to}
        T (\interp{A} \multimap T \interp{B})) \otimes T \interp{A}  
      \\
          &\xstackrel{t}{\to} T(T(\interp{A} \multimap T \interp{B}) \otimes \interp{A})
         \xstackrel{Ts}{\to} TT((\interp{A} \multimap T \interp{B}) \otimes \interp{A})
      \\
      &\xstackrel{\mu}{\to}T((\interp{A} \multimap T \interp{B}) \otimes \interp{A})
        \xstackrel{T\epsilon}{\to}  T T \interp{B}\xstackrel{\mu}{\to} T\interp{B}.
        \end{array}
      \]

    \item Case
      \begin{center}
        \begin{tabular}{l}
          \infer{\Phi \vdash_1 \mathsf{lift}\ M : {!}_\alpha A.}
          {\Phi \vdash_\alpha M : A}
        \end{tabular}
      \end{center}

      By induction hypothesis, we have $\interp{M} : \interp{\Phi} = pX\to \alpha \interp{A}$ for some $X\in \V$.
      By the $\V$-adjunction
      $p \vdash \flat$, we have $\delta \interp{M} : X\to \flat \alpha \interp{A}$. So we define $\interp{\mathsf{lift}\ M} := p\delta{\interp{M}} : pX\to p\flat \alpha \interp{A}$. 
      
    \item Case
      \begin{center}
        \begin{tabular}{l}
          \infer{\Gamma \vdash_{\alpha \& \beta} \mathsf{force}\ M : A.}
          {\Gamma \vdash_\beta M :\ !_\alpha A}
        \end{tabular}
      \end{center}
      We only consider the case where $\alpha = \beta = 0$, the other cases are similar.  
      By induction hypothesis, we have a map $\interp{M} : \interp{\Gamma}\to T p\flat T \interp{A}$.
      Since there is a $\V$-natural transformation $\force : p\flat T \interp{A} \to T \interp{A}$,
      we define $\interp{\force M}$ by
      \[\interp{\Gamma}\xstackrel{\interp{M}}{\to} T p\flat T \interp{A} \xstackrel{T\force}{\to} T T \interp{A} \xstackrel{\mu}{\to} T \interp{A} .\]

    \item Case
      \begin{center}
        \begin{tabular}{l}
          \infer{\Gamma \vdash_\alpha \mathsf{box}\ S \  M : \mathbf{Circ}(S,U).}{\Gamma \vdash_\alpha M : {!}_1 (S \multimap_1 U)}
        \end{tabular}
      \end{center}
      Here we only consider the case $\alpha = 1$. 
      By induction hypothesis, we have $\interp{M} : \interp{\Gamma}\to p\flat (\interp{S} \multimap \interp{U})$.
      We define
      $\interp{\boxt S M}$ by
      \[\interp{\Gamma}\xrightarrow{\interp{M}} p\flat (\interp{S} \multimap \interp{U}) \xstackrel{\boxt}{\to} p\A(\interp{S}, \interp{U}). \]

    \item Case
      \begin{center}
        \begin{tabular}{l}
        \infer[\textit{apply}]{\Phi, \Gamma_1, \Gamma_2 \vdash_{\alpha \& \beta} \mathsf{apply}(M, N) :  U}{\Phi, \Gamma_1 \vdash_\alpha M : \mathbf{Circ}(S,U) & \Phi, \Gamma_2 \vdash_\beta N : S}
        \end{tabular}
      \end{center}
      Here we only consider the case $\alpha = \beta = 0$. 
      By induction hypothesis, we have $\interp{M} : \interp{\Gamma_{1}}\to T p\A(\interp{S}, \interp{U})$
      and  $\interp{N} : \interp{\Gamma_{2}}\to T \interp{S}$. 
      Thus we define
      $\interp{\mathsf{apply}(M, N)}$ by
      \[\def\arraystretch{1.3}
        \begin{array}{l@{~}l}
          \interp{\Phi}\otimes\interp{\Gamma_{1}}\otimes \interp{\Gamma_{2}} 
          &\xrightarrow{\mathsf{dup}\otimes \interp{\Gamma_1}\otimes \interp{\Gamma_2}} \interp{\Phi}\otimes \interp{\Phi}\otimes\interp{\Gamma_{1}}\otimes \interp{\Gamma_{2}}\\
          &\xrightarrow{\interp{M}\otimes \interp{N}} T p\A(\interp{S}, \interp{U})\otimes  T \interp{S}\\
          &\xrightarrow{t} T(T p\A(\interp{S}, \interp{U})\otimes \interp{S})\\
          &\xrightarrow{Ts}  T T(p\A(\interp{S}, \interp{U})\otimes \interp{S})\\
          &\xrightarrow{\mu} T(p\A(\interp{S}, \interp{U})\otimes \interp{S})\\
          &\xrightarrow{T((\mathsf{force}\,\circ\, \unboxt)\,\otimes\,\interp{S})} T((\interp{S}\multimap \interp{U})\otimes \interp{S})\\
          &\xrightarrow{T \epsilon} T \interp{U}.
        \end{array}
      \]
  \end{itemize}
\end{definition}

Our interpretation of the typing rules satisfies the usual semantics substitution theorem. The details of the proof are
in Appendix \ref{app:substition}.
\begin{theorem}[Substitution]
  \label{sem:substitution}
 If $\Phi, \Gamma_{1}, x : A, \Gamma_{2} \vdash_{\alpha} M : B$
  and $\Phi, \Gamma_{3} \vdash_{1} V : A$, then
  \[\interp{[V/x]M} = \interp{M} \circ (\interp{\Phi} \otimes \interp{\Gamma_{1}} \otimes \interp{V} \otimes \interp{\Gamma_{2}})\circ (\mathsf{dup} \otimes \interp{\Gamma_{1}}\otimes \interp{\Gamma_{2}}\otimes \interp{\Gamma_{3}} )
  : \interp{\Phi, \Gamma_{1}, \Gamma_{2}, \Gamma_{3}} \to \alpha \interp{B}.\] 
\end{theorem}

The next two theorems show that values of parameter type are in the image of functor $p$, and that values of simple types are isomorphisms.
\begin{theorem}
  \label{p-value}
  If $\Phi \vdash_{1} V : P$, then $\interp{V} = p f : p X \to p Y$ for some $f : X \to Y \in \V$ such that $\interp{\Phi} = p X, \interp{P} = pY$. 
\end{theorem}

\begin{theorem}
  Suppose $\Sigma \vdash_{1} V : S$, then $\interp{V} : \interp{\Sigma} \to \interp{S}$ is an isomorphism in $\A$.
\end{theorem}
Since the embedding $\psi : \m \hookrightarrow V(\A)$ is fully faithful, $\interp{V} : \interp{\Sigma} \to \interp{S}$ is also an isomorphism in $\m$.

\section{Operational semantics and soundness}
\label{sec:op}
In this section, we will specify an operational semantics for Proto-Quipper-Dyn
and show that it is sound with respect to the $\V$-enriched categorical model $\A$ for dynamic lifting. 

We distinguish two kinds of evaluation in order to model Quipper's two runtimes.
The evaluation rules for circuit generation time will work with morphisms in $\m$, i.e., quantum circuits.
On the other hand, the evaluation rules for circuit execution time will work with morphisms in $\q$, i.e., quantum operations.
Because of the embeddings $\psi : \m \hookrightarrow V(\A)$ and $\phi : \q \hookrightarrow \Kl_{VT}(V(\A))$,
we are able to interpret the configurations for these two runtimes as maps in the $\V$-enriched category $\A$.

\subsection{Operational semantics for circuit generation time}
First of all, we specify the meaning of appending circuits in the category $\m$.

\begin{definition}[Circuit append]
  \label{def:append}
  Suppose $\cC : \Sigma \to \Sigma_{1}, \Sigma_{2}$ and $\dD : \Sigma_{1}' \to \Sigma_{3}$
  are morphisms in $\m$ and there are typing judgments $\Sigma_{1} \vdash_{1} V : S$ and $\Sigma_{1}' \vdash_{1} V' : S$.
  We define $\mathrm{append}(\dD, \cC, V', V)$ to be the following morphism in $\m$.
  \[((\dD\circ \interp{V'}^{-1}\circ \interp{V}) \otimes \interp{\Sigma_{2}})\circ \cC : \Sigma \to \Sigma_{3}, \Sigma_{2} \] 
\end{definition}

Thus $\mathrm{append}(\dD, \cC, V', V)$ is the result of appending the circuit $\dD$ to $\cC$ by connecting the interfaces
$V'$ and $V$.
The following are evaluation rules for circuit generation time, where the underlying states are given by morphisms in $\m$. 

\begin{definition}[Circuit generation time evaluation]
  \label{eval:circ}
  
  \[\small
      \begin{tabular}{llll}
        \\

        \infer{(\cC_1, M N)  \Downarrow (\cC_4, V')}
        {
        \begin{array}{c}
          (\cC_1, M) \Downarrow (\cC_2, \lambda x. M') \\
          (\cC_2, N) \Downarrow (\cC_3, V) \\
          (\cC_3, [V/x]M') \Downarrow (\cC_4, V')
        \end{array}
        }

        &
        \infer[\textit{apply}]{(\cC_1, \mathsf{apply}(M, N))  \Downarrow (\cC', b)}
        {
        \begin{array}{c}
          (\cC_1, M) \Downarrow (\cC_2, (a, \dD, b)) \\
          (\cC_2, N) \Downarrow (\cC_3, V) \\
          \mathrm{append}(\dD, \cC_3, a, V) = \cC'
        \end{array}
        }
                                             
        &
        \infer{(\cC, \force\ M)  \Downarrow (\cC'', V)}
        {
        \begin{array}{c}
          (\cC, M) \Downarrow (\cC', \lift\ M')\\
          (\cC', M') \Downarrow (\cC'', V)
        \end{array}
        }

        \\
        \\
        \infer[\textit{box}]{(\cC, \mathsf{box}\ S\ M) \Downarrow (\cC', (a,\dD,b)) }
        {
        \begin{array}{c}
          (\cC, M) \Downarrow (\cC', \mathsf{lift} \ M') \\
          \mathrm{gen}(S) = a \\
          (\id_S, M'\ a) \Downarrow (\dD, b) 
        \end{array}
        }
        &
        \infer{(\cC, \mathsf{let}\ (x, y) = N\  \mathsf{in}\ M) \Downarrow (\cC'', V)}
        {
        \begin{array}{c}
          (\cC, N) \Downarrow (\cC', (V_1, V_2))\\
          (\cC', [V_1/x, V_2/y]M) \Downarrow (\cC'', V)
        \end{array}
        }
        &
          \infer{(\cC, (M, N)) \Downarrow (\cC'', (V_1, V_2))}
          {
          \begin{array}{c}
            (\cC, M) \Downarrow (\cC', V_1) \\
            (\cC', N) \Downarrow (\cC'', V_2)
          \end{array}
        }          
      \end{tabular}
    \]
\end{definition}
In the rule \textit{box}, we use $\mathrm{gen}(S) = a$ to
mean that the $a$ is a fresh simple term of type $S$.
Note that the evaluation of
$(\cC, M) \Downarrow (\cC', V)$ does not account for
dynamic lifting,
and the underlying states are circuits. So it is the same set of evaluation rules as in \cite{RS2017-pqmodel}.
The evaluation comes with the following notion of configuration. 

\begin{definition}[Well-typed circuit configuration]
  We write $\Sigma \vdash (\cC, M) : A; \Sigma'$ to
  mean there exists $\Sigma''$ such that $\cC : \Sigma \to \Sigma', \Sigma''$ and $\Sigma'' \vdash_1 M : A$. 
\end{definition}

A well-typed circuit configuration requires a typed term with modality $1$, i.e., $\vdash_{1} M : A$.
It is a runtime error if a term with dynamic lifting is encountered when using the evaluation rules in Definition \ref{eval:circ}.
Our type system and the following type preservation theorem ensures
that this can not happen. 

\begin{theorem}
  \label{safe-box}
  If $\Sigma \vdash (\cC, M) : A; \Sigma'$ and $(\cC, M) \Downarrow (\cC', V)$, then $\Sigma \vdash (\cC', V) : A; \Sigma'$. 
\end{theorem}

In the following we define the interpretation $\interp{\cC, M}$ as a map in the $\V$-category $\A$.

\begin{definition}
  Suppose $\Sigma \vdash (\cC, M) : A; \Sigma'$.
  We have maps $\psi \cC : \interp{\Sigma} \to \interp{\Sigma'} \otimes \interp{\Sigma''}$ and $\interp{M} : \interp{\Sigma''} \to \interp{A}$ in $\A$.
  We define $\interp{\cC, M}$ as follows:
  \[\interp{\Sigma} \xstackrel{\psi\cC}{\longrightarrow} \interp{\Sigma'} \otimes \interp{\Sigma''} \xstackrel{\interp{\Sigma'}\otimes \interp{M}}{\longrightarrow} \interp{\Sigma'} \otimes \interp{A}.\]
\end{definition}

The following theorem shows that the evaluation rules for circuit generation time are sound with respect to the categorical model $\A$.
Since in this case dynamic lifting cannot occur, the proof is similar to the one in \cite{RS2017-pqmodel}. 
\begin{theorem}
  \label{circ:sound}
   If $\Sigma \vdash (\cC, M) : A; \Sigma'$ and $(\cC, M) \Downarrow (\cC', V)$,
    then $\interp{\cC, M} = \interp{\cC', V}$.
\end{theorem}

\subsection{Operational semantics for circuit execution time}
Since dynamic lifting requires the ability to access
the states in $\q$, 
we first define the concepts of \textit{state} and
\textit{addresses}.

\begin{definition}[State and addresses]
  For any object $S \in \q$, a \textit{state} is a morphism $Q : I \to S \in \q$. We write $\mathrm{addr}(Q) = \Sigma$ if $\phi(S) = \interp{\Sigma}$, we call $\Sigma$ the \textit{addresses} of $Q$. (Recall that we have, for convenience and without loss of generality, assumed that the interpretation function $\interp{-}$ is one-to-one on label contexts).
\end{definition}

We often write $Q : I \to \Sigma \in \q$ for $Q : I \to S$, where $\phi(S) = \interp{\Sigma}$. The following
\textit{read} operation will be used to define the operational semantics for dynamic lifting.

\begin{definition}[Read operation]
  Suppose $\mathrm{addr}(Q) = \Sigma, \ell : \Bit$ 
  and $Q = p_{1}(Q_{1} \otimes \mathrm{inj}_{1}) + p_{2}(Q_{2}\otimes \mathrm{inj}_{2})$,
  where $\addr{Q_{1}} = \addr{Q_{2}} = \Sigma$ and $p_{1}, p_{2}\in [0, 1]$ and $p_{1} + p_{2} = 1$. 
  We define a formal sum $\mathrm{read}(Q, \ell) = p_{1}(Q_{1}, \mathsf{False}) + p_{2} (Q_{2}, \mathsf{True})$,
  where $\mathsf{False}, \mathsf{True} : \Bool$.
 \end{definition}

 Note that by the last condition in Assumption \ref{ass:mq},
 we know that $Q = p_{1}(Q_{1}\otimes \mathrm{inj}_{1}) + p_{2}(Q_{2}\otimes \mathrm{inj}_{2}) : I \to \Sigma \otimes \Bit$ for some essentially uniquely determined $Q_{1}, Q_{2} : I \to \Sigma$, and $p_{1}, p_{2}\in [0,1]$ such that $p_{1}+p_{2}= 1$. The only time $Q_{i}$ is not uniquely determined is when $p_{i}=0$, but in this case, it will turn out that the $Q_{i}$ does not matter since it corresponds to a branch of computation taken with probability zero. In this case, we can just make some fixed but arbitrary choice for $Q_{i}$.
 So the read operation makes the information of the probabilities $p_{1}, p_{2}$ and the states $Q_{1}, Q_{2}$ available.

 In the following, we define the circuit execution time counterpart of Definition \ref{def:append}.
 It specifies the meaning of updating a quantum state by applying a quantum circuit, where the identity-on-object
 interpretation functor $J : \m \to \q$ is needed for the definition. 
 \begin{definition}
   Suppose $Q : I \to \Sigma_{1}, \Sigma_{2}$ is
   a morphism in $\q$, and $\cC : \Sigma_{1}' \to \Sigma_{3}$ is a morphism in $\m$, 
   and there are typing judgements $\Sigma_{1} \vdash V : S$ and $\Sigma_{1}' \vdash V' : S$.
   We define  $\mathrm{operate}(\cC, Q, V', V)$ to be the following map in $\q$.
   \[ (J(\cC \circ \interp{V'}^{-1} \circ \interp{V}) \otimes \interp{\Sigma_{2}})\circ Q : I \to \Sigma_{3}, \Sigma_{2}\]

 \end{definition}
 
 We now we define the operational semantics for 
 circuit execution time.
 The underlying states of the evaluation are the states in $\q$.
 The evaluation is of the form $(Q, M)  \Downarrow \sum_{i\in [n]} p_{i} (Q_{i}, V_{i})$.
 Its intuitive meaning is that the configuration $(Q, M)$ can be reduced to $(Q_{i}, V_{i})$ with probability $p_{i}$. The
notation $\sum_{i\in [n]} p_{i} (Q_{i}, V_{i})$ is a short hand for the formal sum $p_{1} (Q_{1}, V_{1}) + ... +p_{n} (Q_{n}, V_{n})$,
and we assume $\sum_{i\in [n]}p_{i} = 1$. We write $[n] = \{1, ..., n\}$.  

\begin{definition}[Operational semantics for circuit execution time]
\label{op:dynlift}

  \[\small
    \begin{tabular}{cc}
      \infer{(Q, M N)  \Downarrow \sum_{(i,j,k)\in [n]\times [m]\times [l]} p_i q_{i,j} s_{i,j,k} (Q_{i,j,k}'', V_{i,j,k}')}
      {
      \begin{array}{cc}
        (Q, M) \Downarrow \sum_{i\in [n]} p_i(Q_i, \lambda x. M'_i) \\
        (Q_i, N) \Downarrow \sum_{j\in [m]} q_{i,j} (Q_{i,j}', V_{i,j})\\
        (Q_{i,j}', [V_{i,j}/x]M'_i) \Downarrow \sum_{k\in [l]} s_{i,j, k} (Q_{i,j,k}'', V_{i,j,k}') 
      \end{array}
      }

      &
        \infer{(Q, \force M)  \Downarrow \sum_{(i,j)\in [n]\times [m]}p_i q_{i,j} (Q_{i,j}', V_{i,j})}
      {
      \begin{array}{ll}
        (Q, M) \Downarrow \sum_{i\in [n]}p_i (Q_i, \lift M_i')
        \\ (Q_i, M_i') \Downarrow \sum_{j\in [m]}q_{i,j}(Q_{i,j}', V_{i,j}) 
      \end{array}}
      \\
        \\
        \infer[\textit{apply}]{(Q, \mathsf{apply}(M, N))  \Downarrow \sum_{(i,j)\in [n]\times [m]}p_i q_{i,j} (Q_{i,j}'', b_i)}
        {
        \begin{array}{cc}
          (Q, M) \Downarrow \sum_{i\in [n]}p_i (Q_i, (a_i, \dD_i, b_i)) \\
          (Q_i, N) \Downarrow \sum_{j\in [m]}q_{i,j}(Q_{i,j}', V_{i,j})\\
          \mathrm{operate}(\dD_i, Q_{i,j}', a_{i}, V_{i,j}) = Q_{i,j}''
        \end{array}
      }
      &
      \infer[\textit{box}]{(Q, \mathsf{box}\ S\ M) \Downarrow \sum_{i\in [n]}p_i(Q_i, (a,\dD_i,b_i)) }
      {
      \begin{array}{cc}
        (Q, M) \Downarrow \sum_{i\in [n]}p_i(Q_i, \mathsf{lift} \ M_i') \\
        \mathrm{gen}(S) = a \\
        (\id_S, M_i'\ a) \Downarrow (\dD_i, b_i)
      \end{array}
      }
    \end{tabular}
    \]
    \[
    \begin{tabular}{c}
      \infer[\textit{dynlift}]{(Q, \Dyn M) \Downarrow \sum_{(i, j)\in [n] \times [2]} p_i q_{i,j} (Q_{i,j}', a_{i,j})}
      {
      \begin{array}{cc}
        (Q, M) \Downarrow \sum_{i\in [n]} p_i (Q_i, \ell_i) \\
        \mathrm{read}(Q_i, \ell_i) = q_{i,1}(Q_{i,1}', a_{i,1}) + q_{i,2}(Q_{i,2}', a_{i,2})
      \end{array}
      }
      \\
      \\
      \infer{(Q, \mathsf{let}\ (x, y) = N\  \mathsf{in}\ M) \Downarrow \sum_{(i, j)\in [n]\times [m]}(Q''_{i,j}, V''_{i, j})}{(Q, N) \Downarrow \sum_{i\in [n]}p_i(Q_i', (V_{i}, V'_{i}) ) & (Q'_i, [V_{i}/x, V_{i}'/y]M) \Downarrow \sum_{j\in m}(Q_{i,j}'', V''_{i, j})}          
      \\
      \\
      \infer{(Q, (M, N)) \Downarrow \sum_{i,j\in [n]\times [m]} (Q_{i,j}', (V_i, V_{i,j}'))}
          {
          \begin{array}{c}
            (Q, M) \Downarrow \sum_{i\in [n]}p_{i}(Q_{i}, V_i) \\
            (Q_{i}, N) \Downarrow \sum_{j\in [m]}q_{i,j}(Q_{i,j}', V_{i,j}')
          \end{array}
        }          
    \end{tabular}
  \] 
\end{definition}

In the \textit{apply} rule, we use $\textit{operate}$ instead of $\textit{append}$, which allows a quantum circuit to be applied as
a quantum operation. In the \textit{dynlift} rule, for each $(Q_{i}, \ell_{i})$, we apply the operation
$\mathrm{read}(Q_{i}, \ell_{i})$, which gives rise to two possible outcomes $(Q'_{i,1}, a_{i,1}), (Q'_{i,2}, a_{i, 2})$ with probabilities $q_{i,1}, q_{i,2}$, where $a_{i,1} , a_{i,2} : \Bool$ and $a_{i,1} \not = a_{i, 2}$. This is the only rule that gives rise to probabilistic results in the evaluation. In the \textit{box} rule, the evaluation of
$(\id_{S}, M_{i}'a)$ uses the rules defined in Definition \ref{eval:circ}, so it is
performed at circuit generation time. 

We now define a well-typed configuration for evaluating a term under a quantum state. 
\begin{definition}[Well-typed configuration]
  We write $\vdash_\alpha (Q, M)  : A; \Sigma'$ to mean there exists $\Sigma''$ such that $\Sigma'' \vdash_\alpha M : A$, and $\mathrm{addr}(Q) = \Sigma'', \Sigma'$.
\end{definition}
Since the evaluation rules in Definition \ref{op:dynlift} account
for dynamic lifting, the  above configuration allows the term $M$ to have modality $0$. 
The operational semantics defined in Definition \ref{op:dynlift} is type-safe in the following sense. 

\begin{theorem}
  \label{thm:preserved0}
  If $\vdash_\alpha (Q, M) : A; \Sigma'$ and $(Q, M) \Downarrow \sum_{i\in [n]}p_{i} (Q_{i}, V_{i})$,
  then $\vdash_1 (Q_{i}, V_{i}) : A; \Sigma'$ for all $i\in [n]$.
\end{theorem}

\begin{theorem}
  \label{thm:deterministic}
  If $\vdash_1 (Q, M) : A; \Sigma'$ and $(Q, M) \Downarrow \sum_{i\in
    [n]}p_{i} (Q_{i}, V_{i})$, then $n=1$. In other words, we actually
  have $(Q, M) \Downarrow (Q',V)$.
\end{theorem}

In the following we interpret a well-typed configuration $\vdash_{\alpha} (Q, M) : A; \Sigma'$
as a map in the Kleisli category $\Kl_{T}(\A)$.

\begin{definition}
  Suppose $ \vdash_{\alpha} (Q, M) : A; \Sigma'$.
  We have $\phi Q : I \to T(\interp{\Sigma_{1}'} \otimes \interp{\Sigma''}\otimes \interp{\Sigma_{2}'})$ and $\interp{M} : \interp{\Sigma''} \to \alpha \interp{A}$ in $\A$.
  We define $\interp{Q, M}$ by:
  \begin{itemize}
  \item If $\alpha = 1$, then
    \[I \xstackrel{\phi Q}{\longrightarrow} T(\interp{\Sigma_{1}'} \otimes \interp{\Sigma''} \otimes \interp{\Sigma_{2}'}) \xstackrel{T(\interp{\Sigma'_{1}} \otimes \interp{M} \otimes \interp{\Sigma'_{2}})}{\longrightarrow} T(\interp{\Sigma'_{1}}\otimes \interp{A} \otimes \interp{\Sigma_{2}'}).\]
    
  \item If $\alpha = 0$, then
    \[I \xstackrel{\phi Q}{\longrightarrow} T(\interp{\Sigma'_{1}} \otimes \interp{\Sigma''} \otimes \interp{\Sigma'_{2}}) \xstackrel{T(\interp{\Sigma'_{1}} \otimes \interp{M} \otimes \interp{\Sigma'_{2}})}{\longrightarrow} T(\interp{\Sigma'_{1}} \otimes T\interp{A} \otimes \interp{\Sigma'_{2}})\]
    \[\xstackrel{T(t \otimes \interp{\Sigma'_{2}})}{\to} T(T(\interp{\Sigma'_{1}} \otimes \interp{A}) \otimes \interp{\Sigma'_{2}} ) \xstackrel{Ts}{\to} TT(\interp{\Sigma'_{1}} \otimes \interp{A} \otimes \interp{\Sigma'_{2}} )\xstackrel{\mu}{\to}T(\interp{\Sigma'_{1}} \otimes \interp{A} \otimes \interp{\Sigma'_{2}}).\]
  \end{itemize}
\end{definition}

The following theorem shows that the operational semantics in Definition \ref{op:dynlift} is sound
with respect to the semantic model $\A$.

\begin{theorem}[Soundness]
  \label{thm:soundness}
  If $\vdash_{\alpha} (Q, M) : A; \Sigma'$, and $(Q, M) \Downarrow \sum_{i\in [n]}p_{i} (Q_{i}, V_{i})$,
    then
    \[\interp{Q, M} = \sum_{i\in [n]}p_{i} \interp{Q_{i}, V_{i}} : I \to T(\interp{A}\otimes \interp{\Sigma'}).\]
\end{theorem}

\begin{proof}[Proof sketch]
  The proof is by induction on the evaluation rules.  Here we focus
  on the case for dynamic lifting. Please see Appendix~\ref{app:sound}
  for the proofs of the other cases.
  
  Suppose $\mathrm{addr}(Q) = \Sigma'', \Sigma'$, and $\Sigma'' \vdash_{0} M : A$, 
  and
  \[
    \infer{\Sigma'' \vdash_0 \Dyn M  : \Bool}
    {\Sigma'' \vdash_1 M  : \Bit}.
  \]
  Consider the following. 
  \[\infer{(Q, \Dyn M)\Downarrow q_{1}(Q_{1}', \mathsf{False}) + q_{2}(Q_{2}', \mathsf{True})}
    {
      \begin{array}{c}
        (Q,  M) \Downarrow (Q', \ell)\\
        \mathrm{read}(Q', \ell) = q_{1}(Q_{1}', \mathsf{False}) + q_{2}(Q_{2}', \mathsf{True})
      \end{array}
    }\]

  Since $\mathrm{read}(Q', \ell) = q_{1}(Q_{1}', \mathsf{False}) + q_{2}(Q_{2}', \mathsf{True})$ implies
  that $Q' = q_{1}(Q_{1}' \otimes \mathrm{inj}_{1}) + q_{2}(Q_{2}' \otimes \mathrm{inj}_{2})$ in $\q$, we have
  the following in $\A$.
  \[\phi Q' = q_{1}(\mu \circ T t \circ s \circ (\phi Q_{1}'\otimes \phi(\mathrm{inj}_{1}))) + q_{2} (\mu \circ T t \circ s \circ (\phi Q_{2}'\otimes \phi(\mathrm{inj}_{2}))),\]
  where $\phi Q' : I \to T(\Bit \otimes \interp{\Sigma'})$, and
  $\phi Q_{1}', \phi Q_{2}' : I \to T \interp{\Sigma'}$, and
  $\phi(\mathrm{inj}_{1}), \phi(\mathrm{inj}_{2}) : I \to T\Bit$.
  Note that by condition (\ref{embeddings}), we have $\phi(\mathrm{inj}_{1}) = \eta \circ \mathsf{init}\circ \interp{\mathsf{False}}$ and $\phi(\mathrm{inj}_{2}) = \eta \circ \mathrm{init}\circ \interp{\mathsf{True}}$. 
  We need to show that
  \[\interp{Q, \Dyn M} = q_{1}(T(\interp{\mathsf{False}}\otimes {\interp{\Sigma'}})\circ \phi Q_{1}') + q_{2}(
    T(\interp{\mathsf{True}}\otimes {\interp{\Sigma'}})\circ \phi Q_{2}').\]
  By induction hypothesis, we have $\interp{Q, M} = \interp{Q', \ell}$, i.e., $T(\interp{M}\otimes {\interp{\Sigma'}}) \circ \phi Q = \phi Q'$.
  Thus
  \[\interp{Q, \Dyn M} = \mu \circ T s \circ T((\Dyn \circ \interp{M}) \otimes {\interp{\Sigma'}})\circ \phi Q\]
  \[ = \mu \circ T s \circ T(\Dyn \otimes {\interp{\Sigma'}}  ) \circ T(\interp{M} \otimes {\interp{\Sigma'}})\circ \phi Q\]
  \[= \mu \circ T s \circ T(\Dyn \otimes {\interp{\Sigma'}}) \circ \phi Q' \]
  \[= \mu \circ T s \circ T(\Dyn \otimes {\interp{\Sigma'}}) \]
  \[\circ (q_{1}(\mu \circ T t \circ s \circ (\phi Q_{1}' \otimes \phi(\mathrm{inj}_{1}) )) + q_{2} (\mu \circ T t \circ s \circ (\phi Q_{2}'\otimes \phi(\mathrm{inj}_{2}))))\]
  \[= q_{1} (\mu \circ T s \circ T(\Dyn \otimes {\interp{\Sigma'}}) \circ \mu \circ T t \circ s \circ (\phi Q_{1}'\otimes \phi(\mathrm{inj}_{1}))) \]
  \[+ q_{2} (\mu \circ T \sigma \circ T(\Dyn \otimes {\interp{\Sigma'}}) \circ \mu \circ T t \circ \sigma \circ ( \phi Q_{2}'\otimes \phi(\mathrm{inj}_{2}))).\]

  We just need to show
  \[ T(\interp{\mathrm{False}}\otimes {\interp{\Sigma'}})\circ \phi Q_{1}' = \mu \circ Ts \circ T(\Dyn \otimes {\interp{\Sigma'}}) \circ \mu \circ T t \circ s \circ (\phi Q_{1}'\otimes \phi(\mathrm{inj}_{1}))\]
  \[= \mu \circ Ts \circ T(\Dyn \otimes {\interp{\Sigma'}}) \circ \mu \circ T t \circ s \circ (\phi Q_{1}'\otimes (\eta \circ \mathsf{init}\circ \interp{\mathsf{False}}))\]
  This is true because of the commutative diagram in Figure \ref{fig:diag1}.
  \begin{figure}
    \centering
    \[
      \begin{tikzcd}
        I \arrow[d, "\phi Q_{1}'"]&&&& \\
        T\interp{\Sigma'} \arrow[d, "\lambda^{-1}"] \arrow[dr, "T\lambda^{-1}"] & & & & \\
        I \otimes T\interp{\Sigma'} \arrow[r, "t"]
        \arrow[d, "\interp{\mathsf{False}} \otimes T\interp{\Sigma'}"]& T(I \otimes \interp{\Sigma'})
        \arrow[d, "T(\interp{\mathsf{False}} \otimes \interp{\Sigma'})"]&&& \\
        \Bool \otimes T\interp{\Sigma'}  \arrow[r, "t"]
        \arrow[d, "\mathrm{init} \otimes T\interp{\Sigma'}"]
        & T( \Bool \otimes \interp{\Sigma'})
        \arrow[rr, "\id", bend left = 15]
        \arrow[d, "T(\mathrm{init} \otimes \interp{\Sigma'})"]
        \arrow[dr, "T(\eta \otimes \interp{\Sigma'})", bend left = 5]
        \arrow[drr, "\eta", bend left = 15]
        && T(\Bool \otimes \interp{\Sigma'} ) & \\
        \Bit \otimes T\interp{\Sigma'} 
        \arrow[r, "t"]
        \arrow[d, "\eta \otimes T\interp{\Sigma'}", swap]
        & T(\Bit \otimes \interp{\Sigma'} )
        \arrow[ddl, "T(\eta \otimes \interp{\Sigma'})", near start, swap]
        \arrow[dddl, "\eta", bend left = 10]
        \arrow[r, "T( \Dyn\otimes \interp{\Sigma'})", swap, bend right = 10]
        &T(T\Bool \otimes \interp{\Sigma'} ) \arrow[r, "T s"]
        & TT(\Bool\otimes \interp{\Sigma'}) \arrow[u, "\mu"]& \\
        T \Bit \otimes T\interp{\Sigma'}  \arrow[d, "t"]& & & & \\
        T(T \Bit \otimes \interp{\Sigma'}) \arrow[d, "Ts"] &&&& \\
        TT(\Bit \otimes \interp{\Sigma'} ) \arrow[r, "\mu"]& T(  \Bit \otimes \interp{\Sigma'}) \arrow[uuu, "\id"]&&&
      \end{tikzcd}
    \]
    
    \caption{A commutative diagram from the proof of Theorem~\ref{thm:soundness}}
    \label{fig:diag1}
  \end{figure}
\end{proof}

\begin{remark}
In practice, a closed term $M$ is always evaluated with the initial configuration $(\id_{I}, M)$, where $\id_{I} : I \to I$ is a state
in $\q$. When $M$ has modality $0$, we would need access to a quantum computer/simulator
in order to evaluate $(\id_{I}, M)$ and each run of $(\id_{I}, M)$ could give a different
value. When $M$
has modality $1$, the evaluation of $(\id_{I}, M)$ is deterministic, i.e.,
the top-level quantum state is updated in a deterministic fashion. In
this case, instead of performing the quantum operations, we could also
just generate a list of gates, which can be done entirely in a
classical computer.
\end{remark}

\section{Dynamic lifting in Proto-Quipper}
\label{sec:app}

While typing judgments and certain types ($A\multimap_{\alpha}B$
and $!_{\alpha}A$) are annotated with a modality, information about this modality is meant to be hidden from the programmer unless an error occurs. For example, if one attempt to box a function which uses dynamic lifting, the type checker will raise a modality error. As a result, such programming errors are caught at compile time in Proto-Quipper-Dyn, whereas they are only caught at runtime in Quipper.

The modality inference can readily be integrated into bi-directional type checking, which uses a pair of recursively defined functions for type checking and type inference \cite{pierce2000local}. To work with modalities, the type checking function not only takes a term and a type as inputs, but also the current modality of the typing judgment.
For example, when checking a term $\lambda x .M $ against a type $A \multimap_{\alpha} B$ with current modality $\beta$,
the type checking function first ensures that the current modality $\beta$ is $1$, then extends the current
typing environment with $x : A$ and recursively checks the term $M$ against the type $B$, with the
modality $\alpha$. The type inference function takes a term as input and outputs the inferred type as well as the inferred modality.
For example, when inferring the type for a term $M N$, the type inference function first
infers a type $A \multimap_{\alpha}B$ and a modality $\beta$ for $M$, and then it infers a type $A$ and a modality
$\gamma$ for the term $N$, so the type inference function will return the type $B$ and the inferred modality $\alpha \& \beta \& \gamma$.

We now discuss several Proto-Quipper-Dyn programs that make use of
dynamic lifting. An experimental implementation of Proto-Quipper-Dyn
is available from {\url{https://gitlab.com/frank-peng-fu/dpq-remake}}, and the programs in
Listings~\ref{lst:alice-bob-circ}--\ref{lst:rus} have been tested with
that implementation. Note that several of the following example
programs make use of recursion. While we do not formally treat
recursion in this paper, it is included in the prototype
implementation.
 
\subsection{Quantum teleportation}
The following circuit implements a one-qubit quantum teleportation protocol. 
\[
\includegraphics[width=0.5\textwidth]{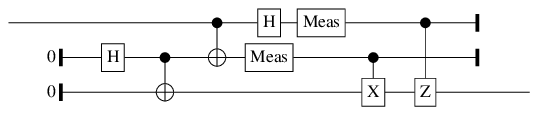}
\]
This circuit is generated by the following Proto-Quipper-Dyn programs.
\vspace{-2em}

\begin{center}
\begin{minipage}[t]{0.45\textwidth}
  \begin{lstlisting}[frame=single,caption={Alice and Bob circuits}, label={lst:alice-bob-circ}]
alice1 : !(Qubit -> Qubit -> Bit * Bit)
alice1 a q =
  let (a, q) = CNot a q
      q = H q
  in (Meas a, Meas q)

bob1 : !(Qubit -> Bit -> Bit -> Qubit)
bob1 q x y =
  let (q, x) = C_X q x 
      (q, y) = C_Z q y
      _ = Discard x
      _ = Discard y
  in q
\end{lstlisting}
\end{minipage}\hspace{0.05\textwidth}
\begin{minipage}[t]{.45\textwidth}
  \begin{lstlisting}[caption={Teleportation circuit},frame=single,label={lst:tele-circ}]
bell00 : !(Unit -> Qubit * Qubit)
bell00 u =  
  let a = Init0 ()
      b = Init0 ()
  in CNot b (H a)
      
tele1 : !(Qubit -> Qubit)
tele1 q =
  let (b, a) = bell00 ()
      (x, y) = alice1 a q
      z = bob1 b x y
  in z

boxTele : Circ(Qubit, Qubit)
boxTele = box Qubit tele1      
  \end{lstlisting}
\end{minipage}
\end{center}

As can be seen in Listings \ref{lst:alice-bob-circ} and \ref{lst:tele-circ}, the modality information is not visible to the programmer. Because the programs in Listings \ref{lst:alice-bob-circ} and \ref{lst:tele-circ} do not use dynamic lifting, the
modalities in the fully annotated types are all $1$. For example, the
fully annotated type of \textsf{tele1} is $!_{1}(\Qubit \multimap_{1} \Qubit)$. We can therefore box \textsf{tele1} into a quantum circuit. The evaluation of \textsf{boxTele} occurs on a classical computer and generates the circuit diagram above. 

For comparison, let us consider the following Proto-Quipper-Dyn programs that implement quantum teleportation using dynamic lifting.

\begin{center}
\begin{minipage}[t]{.45\textwidth}
  \begin{lstlisting}[frame=single,caption={Alice and Bob functions}, label={lst:alice-bob}]
alice2 : !(Qubit -> Qubit -> Bool * Bool)
alice2 a q =
  let (a, q) = CNot a q
      q = H q
  in (dynlift (Meas a), dynlift (Meas q))
         
bob2 : !(Qubit -> Bool -> Bool -> Qubit)
bob2 q x y =
  let q = if x then QNot q else q
      q = if y then ZGate q else q
  in q


  
\end{lstlisting}
\end{minipage}\hspace{0.05\textwidth}
\begin{minipage}[t]{.45\textwidth}
  \begin{lstlisting}[caption={Teleportation function},frame=single,label={lst:tele}]
tele2 : !(Qubit -> Qubit)
tele2 q =
  let (b, a) = bell00 ()
      (x, y) = alice2 a q
      z = bob2 b x y
  in z

-- The following will raise an error  
boxAttempt : Circ(Qubit, Qubit)
boxAttempt = box Qubit tele2

test : Bool
test =
  dynlift (Meas (tele2 (Init0 ())))
\end{lstlisting}
\end{minipage}
\end{center}

As before, the code in Listings \ref{lst:alice-bob} and \ref{lst:tele} contains no modality annotations. In the \textsf{alice2}
function, dynamic lifting is used right after the measurement gate $\Meas : \Qubit \to \Bit$. Accordingly, the fully annotated type 
of \textsf{alice2} is $!_{1}(\Qubit \multimap_{1} \Qubit \multimap_{0} \Bool * \Bool)$.
The \textsf{bob2} function then uses if-then-else expressions to decide whether to apply
the gates \textsf{QNot} and \textsf{ZGate}, rather than applying the bit-controlled gates
\textsf{C\_X} and \textsf{C\_Z}, as in the \textsf{bob1} function in Listing~\ref{lst:alice-bob-circ}. 

The \textsf{tele2} function calls the \textsf{bob2} function with the booleans provided by the \textsf{alice2} function. Hence, the \textsf{tele2} function implicitly uses dynamic lifting. Its fully annotated type is $!_{1}(\Qubit \multimap_{0} \Qubit)$.
Because of the modality inference, the type checker will issue a typing error for the \textsf{boxAttempt} function. According to the typing rule for \textit{box}, the \textsf{box Qubit} function requires an argument of type $!_{1}(\Qubit \multimap_{1} \Qubit)$, which is distinct from the type of \textsf{tele2}. This error is sensible because the \textsf{tele2} function does not correspond to a circuit.

The \textsf{test} function applies \textsf{tele2} to an input qubit in the $\ket{0}$ state. The output value of \textsf{test} should then be \textsf{False} with probability $1$. Note that the evaluation of \textsf{test} requires access to a quantum computer or a simulator. 

\subsection{Magic state distillation}
\label{sec:magic}

\textit{Magic states} are quantum states that can be used, in conjunction with Clifford gates, to perform universal quantum computing fault tolerantly \cite{bravyi2005universal}. For example, there is a standard method to implement a $T$ gate using the magic state $(\ket{0} + e^{\frac{\pi i}{4}}\ket{1})/\sqrt{2}$, along with Clifford gates and measurements. This enables the application of any operation from the Clifford+$T$ gate set, a well-known universal set of quantum gates \cite{nielsen2002quantum}.

The process of producing a magic state such as $(\ket{0} + e^{\frac{\pi i}{4}}\ket{1})/\sqrt{2}$ from several imperfect states is called \textit{magic state distillation} \cite{bravyi2005universal}. In order to distill a magic state $\ket{M}$, one first prepares several qubits in a state that approximates $\ket{M}$ up to an error rate $\epsilon$. A carefully designed quantum circuit is then applied to these qubits and some of them are measured. If all of the measurement results are 0, then the remaining qubits are guaranteed to be in a state that approximates $\ket{M}$ up to an improved error rate $\epsilon' < \epsilon$. If any one of the measurement results is 1, then all of the qubits are discarded and the entire process is restarted. In practice, several rounds of distillation are required to obtain a state that approximates $\ket{M}$ up to an acceptable error rate.

\begin{figure}
\begin{center}
\begin{minipage}{0.95\textwidth}
\begin{lstlisting}[frame=single,caption={Bravyi and Kitaev's algorithm}, label={lst:magic}]
distill : ! (Qubit * Qubit * Qubit *  Qubit * Qubit -> Maybe Qubit)
distill input =
  let (a1, a2, a3, a4, a5) = fiveQubits input
      a1' = dynlift (Meas a1)
      a2' = dynlift (Meas a2)
      a3' = dynlift (Meas a3)
      a4' = dynlift (Meas a4)
  in if a1' || a2' || a3' || a4'
     then let a = dynlift (Meas a5) in Nothing
     else Just a5

distillation : ! (Nat -> Qubit)
distillation n =
  case n of
    Z -> prepMixedState ()
    S n' -> 
     let q1 = distillation n' 
         q2 = distillation n'
         q3 = distillation n'
         q4 = distillation n'
         q5 = distillation n'
     in
     case distill (q1, q2, q3, q4, q5) of
       Nothing -> distillation n
       Just q -> q
\end{lstlisting}
\end{minipage}
\end{center}
\end{figure}

A Proto-Quipper-Dyn implementation of Bravyi and Kitaev's distillation algorithm is given in Listing \ref{lst:magic}. 
In the \textsf{distill} function, we first apply a five-qubit error correction circuit \textsf{fiveQubits} to the inputs, then measure the qubits and, through dynamic lifting, promote the resulting bits to booleans.
If all of the booleans are \textsf{False}, the distillation was successful and we return the remaining qubit. Otherwise the distillation failed, so we discard the unmeasured qubit and return nothing. 
Dynamic lifting is essential for defining the \textsf{distill} function because in this case the if-then-else expression cannot be implemented as a circuit. 

The \textsf{distillation} function performs $n$ rounds of magic state distillation. The function \textsf{prepMixedState} prepares an initial imperfect state. The \textsf{distillation} function is a recursive function that assumes five successful distillations from the previous round and then applies the \textsf{distill} function to the resulting qubits. If that function returns a qubit, the $n$-th round of distillation was successful, otherwise it will restart the whole process. 

\subsection{Repeat-Until-Success}

The \emph{repeat-until-success} paradigm provides a technique to apply a unitary that cannot be implemented exactly, at the cost of potentially running the same circuit multiple times. In order to apply a non-Clifford+T gate $N$ to a target qubit $\ket{\phi}$, one first initializes several ancillary qubits before applying a well-chosen Clifford+T circuit $C$ to the target and the ancillas and measuring the ancillas. If all of the measurement results are 0, the target qubit is guaranteed to be in the state  $N\ket{\phi}$. Otherwise, a correction is applied to the target to return it to its initial state and the process is repeated.

Consider the following circuit used in \cite{paetznick2013repeat} to illustrate the implementation of the gate $V_{3} = \frac{I + 2iZ}{\sqrt{5}}$ using the repeat-until-success method. 
\[
\includegraphics[width=0.5\textwidth]{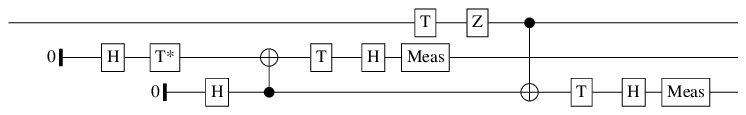}
\]
The top wire is the target qubit, while the wires below it are the ancillas. We apply a sequence of gates ($H$, $H$, $T^{*}$, $CNOT$, $T$, and $H$) to the ancillas before measuring the first ancilla. If, as we assume here, the measurement result is 0, then we apply a sequence of gates ($T$, $Z$, $CNOT$, $T$, and $H$) to the target qubit and the second ancilla before measuring the second ancilla. Assuming, again, that the measurement result is 0, we then know that the target qubit is in the desired state. Note that the circuit above is not a representation of the entire repeat-until-success protocol. Instead, it is the circuit constructed in the event that both measurement results are 0 (which can be shown to occur with probability 5/8). If the measurements yield different results, the circuit constructed by the repeat-until-success protocol is different. For example, if the result of the second measurement is 1, a $Z$ gate must be applied to the target qubit to return it to its initial state.

Listing~\ref{lst:rus} gives a precise description of the implementation of $V_{3}$ in Proto-Quipper-Dyn.
 
\begin{center}
\begin{minipage}{0.95\textwidth}
\begin{lstlisting}[frame=single,caption={A repeat-until-success example}, label={lst:rus}]
v3 : !(Qubit -> Qubit)
v3 q =
  let a1 = tgate_inv (H (Init0 ()))
      a2 = H (Init0 ())
      (a1, a2) = CNot a1 a2
      a1 = H (TGate a1)
  in if dynlift (Meas a1)
     then
       let _ = Discard (Meas a2)
       in v3 q
     else let q = ZGate (TGate q)
              (a2, q) = CNot a2 q
              a2 = H (TGate a2)
          in if dynlift (Meas a2)
             then v3 (ZGate q)
             else q
\end{lstlisting}
\end{minipage}
\end{center}

Once again, dynamic lifting plays an essential role here. Note that the \textsf{v3} function has type $!_{1}(\Qubit \multimap_{0}\Qubit)$; it is a quantum computation, rather than a quantum circuit.

\section{Conclusion}
\label{sec:conclude}

We have given an axiomatization of an enriched categorical semantics for Proto-Quipper with dynamic lifting.
We defined a type system with a modality
to keep track of functions that use dynamic lifting. The main benefit of our type system is that
it statically ensures that 
the boxing operation can only be applied to a function that
does not use dynamic lifting.
We also gave an operational semantics 
for dynamic lifting. The operational semantics models both circuit generation and circuit execution. 
We also defined an abstract categorical semantics for this language
and proved that the type system and the operational semantics 
are sound with respect to it.
Lastly, we gave some examples of quantum algorithms that rely on
dynamic lifting.

There are many things left to be done. One of them is how to combine
dynamic lifting with dependent types and/or recursion. At this point,
we have a model for dynamic lifting {\cite{FKRS-model-2022}}, but it does
not support dependent types or recursion. We also have a model for
Proto-Quipper with dependent types {\cite{FKS2020-lindep}}, but it does not
support dynamic lifting or recursion. Finally, Lindenhovius et al.\@
have a model for Proto-Quipper with recursion {\cite{BMZ2018}}, but it
does not support dynamic lifting or dependent types. We do not think
that adding recursion to Proto-Quipper-Dyn would create fundamental
difficulties at the level of the syntax, type system, or operational
semantics (although languages with recursion are usually better
handled by small-step operational semantics rather than the big-step
semantics we considered here). On the other hand, finding a concrete
denotational semantics gets more complicated the more features are
included in the programming language, and how to add dependent types
or recursion to a semantics of Proto-Quipper-Dyn is an open problem.

\section*{Acknowledgements}

We thank the referees for their thoughtful comments. This work was
supported by the Natural Sciences and Engineering Research Council of
Canada (NSERC) and by the Air Force Office of Scientific Research
under Award No.\@ FA9550-21-1-0041.

\bibliography{dynlift}

\appendix
\section{Diagrams for strong monoidal $\V$-functors}
\label{app:strong-monoidal}
\begin{definition}
  Suppose $\A, \mathbf{B}$ are $\V$-categories. A $\V$-functor $F : \A \to \mathbf{B}$ is 
  \emph{strong monoidal} if the following hold.
  \begin{itemize}
    
  \item In $\mathbf{B}$, there is an isomorphism $e: I \to FI$ and a
    $\V$-natural isomorphism $m : FA \otimes FB \to F(A\otimes B)$.
  \item For all $A, B, C\in \A$, the following diagrams commute.
    \[
      \begin{tikzcd}
        (FA \otimes FB)\otimes FC \arrow[r, "\alpha"]\arrow[d, "m\otimes FC"] &  FA \otimes (FB\otimes FC) \arrow[d, "FA\otimes m"]\\
        F(A \otimes B)\otimes FC \arrow[d, "m"] & FA \otimes F(B\otimes C)\arrow[d, "m"] \\
        F((A \otimes B)\otimes C)\arrow[r, "F\alpha"] & F(A \otimes (B\otimes C)
      \end{tikzcd}
    \]
    \[
      \begin{tikzcd}
        I \otimes FA \arrow[r, "e\otimes FA"] \arrow[d, "l"] & FI \otimes FA \arrow[d, "m"]\\
        FA & F(I\otimes A) \arrow[l, "F l"]
      \end{tikzcd}
    \]
    \[
      \begin{tikzcd}
         FA \otimes I \arrow[r, "FA\otimes e"] \arrow[d, "r"] &  FA \otimes FI \arrow[d, "m"]\\
        FA & F(A \otimes I) \arrow[l, "F r"]
      \end{tikzcd}
    \]
  \end{itemize}

\end{definition}

\section{Convexity}
\label{app:def-convex}
Let $\runit$ denote the real unit interval. 
\begin{definition}
  A \emph{convexity structure} on a set $X$ is an operation that assigns to all $p,q\in\runit$ with $p+q=1$ and all $x,y\in X$ an element $px+qy\in X$, subject to the following properties. Throughout, we assume $p+q=1$.
  \begin{itemize}
  \item[(a)] $px + qx = x$ for all $x\in X$. 
    
  \item[(b)]  
    $p x + q y = q y + p x$ for all $x, y \in X$.
    
  \item[(c)] $0x + 1y = y$ for all $x, y \in X$.
    
  \item[(d)] $(a+b)(\frac{a}{a+b} x + \frac{b}{a+b} y) +
    (c+d)(\frac{c}{c+d} z + \frac{d}{c+d} w) =
    (a+c)(\frac{a}{a+c} x + \frac{c}{a+c} z) +
    (b+d)(\frac{b}{b+d} y + \frac{d}{b+d} w)$, where
    $a,b,c,d\in\runit$ with $a+b+c+d=1$ and
    all denominators are non-zero.
  \end{itemize}
\end{definition}

\noindent \textbf{Remark.} Property $(d)$ can best be understood by
realizing that both sides of the equation are equal to $ax+by+cz+dw$,
decomposed in two different ways into convex sums of two elements at a
time. In the literature, we sometimes find a different, but equivalent
condition of the form $s(px + qy) + r z = spx +
(qs+r)(\frac{qs}{qs+r}y + \frac{r}{qs+r}z)$. The latter axiom is arguably
shorter, but harder to read.

We often expand the binary $+$ operation to a multi-arity operation,
i.e., $\sum_{i} p_{i}x_{i}$, where $\sum_{i}p_{i} = 1$ and $x_{i} \in X$ for all $i$.

\section{Details of Definition~\ref{def:interpretation}}
\label{app:interpretation}
\begin{definition}
  To each valid typing judgement $\Gamma \vdash_\alpha M : A$, we associate a map $\interp{M} : \interp{\Gamma} \to  \alpha \interp{A}$ in $\A$, called its \emph{interpretation}. Note that $\interp{M}$ here is an abbreviation for $\interp{\Gamma \vdash_\alpha M : A}$. 
  The interpretation is defined by induction on the derivation of
  $\Gamma \vdash_\alpha M : A $. We show a few non-trivial cases. 

  \begin{itemize}
  \item Case
    \begin{center}
      \begin{tabular}{l}
        \\
        \infer
        {\Phi, x : A \vdash_1 x : A.}{}
      \end{tabular}
    \end{center}
    We define $\interp{x}$ as
    \[
      \begin{tikzcd}
        \interp{\Phi}\otimes \interp{A} \arrow[rr, "\mathsf{discard} \otimes \interp{A}"] & & \interp{A}.
      \end{tikzcd}
    \]
    \item Case
      \[
        \infer
        {\Phi, \Gamma_1, \Gamma_2 \vdash_{\alpha_1 \& \alpha_2} (M, N) : A\otimes B.}
        {\Phi, \Gamma_1 \vdash_{\alpha_1} M :  A & \Phi, \Gamma_2 \vdash_{\alpha_2} N : B}
      \]

      Suppose $\alpha_{1} = \alpha_{2} = 0$. The other cases are similar.
      By induction hypothesis, we have $\interp{M} : \interp{\Phi}\otimes \interp{\Gamma_{1}} \to T\interp{A}$ and $\interp{N} : \interp{\Phi}\otimes \interp{\Gamma_{2}}\to T\interp{B}$.
      So we define $\interp{M, N}$ as the following composition
      \[\interp{\Phi}\otimes \interp{\Gamma_{1}}\otimes \interp{\Gamma_{2}} \xstackrel{\mathsf{dup}\otimes \interp{\Gamma_{1}}\otimes \interp{\Gamma_{2}}}{\to} \interp{\Phi}\otimes \interp{\Phi}\otimes \interp{\Gamma_{1}}\otimes \interp{\Gamma_{2}} \xstackrel{\interp{M}\otimes \interp{N}}{\to} T\interp{A}\otimes T\interp{B}\]
      \[\xstackrel{t}{\to} T(T\interp{A}\otimes \interp{B})
        \xstackrel{Ts}{\to} TT(\interp{A}\otimes \interp{B})
        \xstackrel{\mu}{\to} T(\interp{A}\otimes \interp{B}).\]
    \item Case
      \begin{center}
        \begin{tabular}{l}
          \\
          \infer{\Phi \vdash_1 (a, \cC, b) :  \mathbf{Circ}(S, U)}
          {
          \begin{array}{ll}
            \Sigma_1 \vdash_1 a : S
            &
              \Sigma_2 \vdash_1 b : U \\
            \cC : \Sigma_1 \to \Sigma_2
          \end{array}
          }
        \end{tabular}
      \end{center}

      Note that $\interp{a} : \interp{\Sigma_{1}} \to \interp{S}$
      and $\interp{b} : \interp{\Sigma_{2}} \to \interp{U}$ are isomorphisms. Moreover, $\cC : \interp{\Sigma_{1}} \to \interp{\Sigma_{2}}$ is a morphism in $\A$. 
      We write $\theta_{\cC} = \interp{b}\circ \cC \circ \interp{a}^{-1} : \interp{S} \to \interp{U}$. Thus we have $\mathsf{curry}(\theta_{\cC}) : I \to \interp{S}\multimap \interp{U}$, and $\delta ({\mathsf{curry}(\theta_{\cC})}) : 1 \to \flat (\interp{S}\multimap \interp{U})$ for the adjunct mate of $\mathsf{curry}(\theta_{\cC})$. 
      Thus we define $\interp{(a, \cC, b)}$ as
      \[ \interp{\Phi} \xstackrel{\mathsf{discard}}{\to} I = p1 \xstackrel{p\delta({\mathsf{curry}(\theta_{\cC})})}{\longrightarrow} p\flat (\interp{S}\multimap \interp{U}) \xstackrel{\mathsf{box}}{\longrightarrow} p\A(\interp{S}, \interp{U}). \]

    \item Case ``let''. We only consider the following; the other
      cases are similar.
      \[
        \infer[\textit{let}]
        {\Phi, \Gamma_1, \Gamma_2 \vdash_{0} \mathsf{let}\ (x, y) = N\  \mathsf{in}\ M : C}
        {\Phi, \Gamma_1, x : P, y : B \vdash_{0} M :  C &  \Phi, \Gamma_2 \vdash_{0} N : P\otimes B}
      \]

      By induction hypothesis, we have $\interp{M} : \interp{\Phi}\otimes \interp{\Gamma_{1}}\otimes \interp{P} \otimes \interp{B} \to T\interp{C}$
      and $\interp{N} : \interp{\Phi}\otimes \interp{\Gamma_{2}} \to T(\interp{P}\otimes \interp{B})$. Thus we define
      $\interp{\mathsf{let}\ (x, y) = N\  \mathsf{in}\ M}$
      as the following composition.
      \[
        \interp{\Phi}\otimes \interp{\Gamma_{1}}\otimes \interp{\Gamma_{2}} \xstackrel{\mathsf{dup}\otimes \interp{\Gamma_{1}}\otimes \interp{\Gamma_{2}} }{\to} \interp{\Phi}\otimes \interp{\Phi}\otimes \interp{\Gamma_{1}}\otimes \interp{\Gamma_{2}} 
      \]
      \[\xstackrel{\interp{\Phi}\otimes \interp{\Gamma_{1}}\otimes \interp{N}}{\to} \interp{\Phi}\otimes \interp{\Gamma_{1}}\otimes T(\interp{P}\otimes \interp{B})\xstackrel{t}{\to}\]
      \[T(\interp{\Phi}\otimes \interp{\Gamma_{1}}\otimes \interp{P}\otimes \interp{B}) \xstackrel{T\interp{M}}{\to} TT\interp{C}\xstackrel{\mu}{\to} T\interp{C}.\]
  \end{itemize}
\end{definition}

\section{Proof of Theorem \ref{sem:substitution}}
\label{app:substition}
\begin{theorem}[Substitution]
  \noindent If $\Phi, \Gamma_{1}, x : A, \Gamma_{2} \vdash_{\alpha} M : B$
  and $\Phi, \Gamma_{3} \vdash_{1} V : A$, then
  \[\interp{[V/x]M} = \interp{M} \circ (\interp{\Phi} \otimes \interp{\Gamma_{1}} \otimes \interp{V} \otimes \interp{\Gamma_{2}})\circ (\mathsf{dup} \otimes \interp{\Gamma_{1}}\otimes \interp{\Gamma_{2}}\otimes \interp{\Gamma_{3}} )   : \interp{\Phi, \Gamma_{1}, \Gamma_{2}, \Gamma_{3}} \to \alpha \interp{B}.\] 
\end{theorem}
\begin{proof}
  By induction on $\Phi, \Gamma_1, x : A, \Gamma_2 \vdash_\alpha M : B$.
  Here we show a few nontrivial cases.
  \begin{itemize}
    \item Case
      \begin{center}
        \begin{tabular}{l}
          \\
          \infer{\Phi, \Gamma_1, x : A, \Gamma_1'  \vdash_0 \Dyn M  : \Bool.}
          {\Phi, \Gamma_1, x : A, \Gamma_1' \vdash_\alpha M  : \Bit}
        \end{tabular}
      \end{center}

      Suppose $\Phi, \Gamma_{2} \vdash_{1} V : A $ and $\alpha = 0$. By induction hypothesis,
      we have
      \[\interp{[V/x]M} = \interp{M} \circ (\interp{\Phi}\otimes \interp{\Gamma_1} \otimes \interp{V} \otimes \interp{\Gamma_2}) \circ (\mathsf{dup} \otimes \interp{\Gamma_1}\otimes \interp{\Gamma_1'}\otimes \interp{\Gamma_2})\]
      \[:  \interp{\Phi} \otimes \interp{\Gamma_1} \otimes \interp{\Gamma_2} \otimes \interp{\Gamma_1'} \to T \interp{\Bit}.\]

      Since $\Phi, \Gamma_{1}, \Gamma_{2}, \Gamma_{1}' \vdash_{0} \Dyn [V/x]M : \Bool$, we have
      \[\interp{\Dyn [V/x]M} =  \mu \circ T \Dyn \circ \interp{ [V/x]M}\]
      \[
      =  (\mu \circ T \Dyn \circ \interp{M}) \circ (\interp{\Phi}\otimes \interp{\Gamma_1} \otimes \interp{V} \otimes \interp{\Gamma_2}) \circ (\mathsf{dup} \otimes \interp{\Gamma_1}\otimes \interp{\Gamma_1'}\otimes \interp{\Gamma_2})\]
      \[: \interp{\Phi} \otimes \interp{\Gamma_{1}} \otimes \interp{\Gamma_{2}} \otimes \interp{\Gamma_{1}'} \to T  \interp{\Bool}.\]

    \item Case
      \begin{center}
        \begin{tabular}{l}
          \\
          \infer{\Phi, x: P, \Phi' \vdash_1 \mathsf{lift}\ M : {!}_\alpha A.}
          {\Phi, x: P, \Phi' \vdash_\alpha M : A}
        \end{tabular}
      \end{center}
      By Lemma \ref{lem:value}, we have $\Phi \vdash_{1} V : P$.
      Since $\Phi, \Phi' \vdash_\alpha [V/x] M : A$, by induction hypothesis,
      we have
      \[\interp{[V/x]M} = \interp{M} \circ (\interp{\Phi} \otimes \interp{V} \otimes \interp{\Phi'}) \circ (\mathsf{dup}\otimes \interp{\Phi'}) :  \interp{\Phi} \otimes \interp{\Phi'} \to \alpha \interp{A}.\]
      Since $\Phi, \Phi' \vdash_1 \lift [V/x] M : {!}_{\alpha}A$, we have
      \[\interp{\lift [V/x]M} = p \delta \interp{[V/x] M} :  \interp{\Phi} \otimes \interp{\Phi'} \to p\flat \alpha \interp{A}.\]
      Note that $\delta \interp{[V/x] M}$ is the unique morphism such
      that $\force \circ p\delta \interp{[V/x] M} = \interp{[V/x] M}$.
      On the other hand, we have
      \[ \force \circ p \delta \interp{M} \circ (\interp{\Phi} \otimes \interp{V} \otimes \interp{\Phi'}) \circ (\mathsf{dup}\otimes \interp{\Phi'}) \]
      \[= \interp{M} \circ (\interp{\Phi} \otimes \interp{V} \otimes \interp{\Phi'}) \circ (\mathsf{dup}\otimes \interp{\Phi'}) = \interp{[V/x]M}.\]
      Thus
      \[\interp{\lift [V/x]M} = p \delta \interp{[V/x] M} = p \delta \interp{M} \circ (\interp{\Phi} \otimes \interp{V} \otimes \interp{\Phi'})\circ (\mathsf{dup}\otimes \interp{\Phi'})\]
      \[ = \interp{\lift M} \circ (\interp{\Phi} \otimes \interp{V} \otimes \interp{\Phi'})\circ (\mathsf{dup}\otimes \interp{\Phi'}).\]

    \item Case
      \begin{center}
        \begin{tabular}{l}
          \\
          \infer{\Gamma_1 , x : C, \Gamma_1'\vdash_{1} \lambda y . M : A \multimap_\alpha B.}
          {\Gamma_1, x : C, \Gamma_1', y : A \vdash_{\alpha} M : B} %
        \end{tabular}
      \end{center}
      Suppose $\Gamma_{2} \vdash_{1} V : C$. By induction hypothesis, we have
      \[ \interp{[V/x]M} = \interp{M} \circ (\interp{\Gamma_{1}} \otimes \interp{V} \otimes \interp{\Gamma_{1}', y : A}) \]
      \[ : \interp{\Gamma_{1}} \otimes \interp{\Gamma_{2}}\otimes \interp{\Gamma_{1}', y : A} \to \alpha \interp{B}.\]
      Since $\Gamma_1 , \Gamma_{2}, \Gamma_1'\vdash_{1} \lambda y .[V/x] M : A \multimap_\alpha B$, we have
      \[\interp{\lambda y. [V/x]M} = \mathsf{curry}{\interp{[V/x]M}} :
      \interp{\Gamma_{1}} \otimes \interp{\Gamma_{2}}\otimes
      \interp{\Gamma_{1}'} \to \interp{A}\multimap \alpha \interp{B},\]
      which is the unique morphism such that
      \[ \epsilon \circ (\mathsf{curry}{\interp{[V/x]M}} \otimes {\interp{A}}) =  \interp{[V/x]M}.
      \]
      On the other hand, we have
      \[ \epsilon \circ ((\mathsf{curry}{\interp{M}} \circ (\interp{\Gamma_{1}} \otimes \interp{V} \otimes \interp{\Gamma_{1}', y : A}))  \otimes {\interp{A}})
      \]
      \[ = \epsilon \circ (\mathsf{curry}{\interp{M}} \otimes {\interp{A}}) \circ (\interp{\Gamma_{1}} \otimes \interp{V} \otimes \interp{\Gamma_{1}', y : A})
      \]
      \[ = \interp{M} \circ (\interp{\Gamma_{1}} \otimes \interp{V} \otimes \interp{\Gamma_{1}', y : A}) = \interp{[V/x]M}.\]
      Thus $\interp{\lambda y. [V/x]M} = \mathsf{curry}{\interp{M}} \circ (\interp{\Gamma_{1}} \otimes \interp{V} \otimes \interp{\Gamma_{1}', y : A}) = \interp{\lambda y.M} \circ (\interp{\Gamma_{1}} \otimes \interp{V} \otimes \interp{\Gamma_{1}', y : A})$.

    \item Case
      \begin{center}
        \begin{tabular}{l}
          \infer {\Phi, \Gamma_1, x : C, \Gamma_1', \Gamma_2 \vdash_{\alpha_1 \& \alpha_2 \& \beta} M N : B}
          {\Phi,\Gamma_1, x : C, \Gamma_1' \vdash_{\alpha_1} M :  A \multimap_\beta B & \Phi, \Gamma_2 \vdash_{\alpha_2} N : A}
        \end{tabular}
      \end{center}

      Suppose $\Phi, \Gamma_{3} \vdash_{1} V : C$ and $\alpha_{1} = \alpha_{2} = \beta = 0$. By induction hypothesis, we have
      \[ \interp{[V/x] M} = \interp{M} \circ (\interp{\Phi} \otimes \interp{\Gamma_{1}} \otimes \interp{V} \otimes \interp{\Gamma_{1}'}) \circ (\mathsf{dup}\otimes \interp{\Gamma_{1}} \otimes \interp{\Gamma_{1}'}\otimes \interp{\Gamma_{3}})\]
      \[: \interp{\Phi} \otimes \interp{\Gamma_{1}} \otimes \interp{\Gamma_{3}} \otimes \interp{\Gamma_{1}'} \to T (\interp{A} \multimap T\interp{B}).\]

      Moreover, \[ \interp{[V/x]MN} = \mu \circ T\epsilon  \circ \mu \circ Ts \circ t \circ (\interp{[V/x]M} \otimes \interp{N}) = \mu \circ T\epsilon  \circ \mu \circ Ts \circ t \circ\]
      \[ ((\interp{M} \circ (\interp{\Phi} \otimes \interp{\Gamma_{1}} \otimes \interp{V} \otimes \interp{\Gamma_{1}'}) \circ (\mathsf{dup}\otimes \interp{\Gamma_{1}} \otimes \interp{\Gamma_{1}'}\otimes \interp{\Gamma_{3}})) \otimes \interp{N}).\]
      On the other hand, \[\interp{M N} \circ (\interp{\Phi} \otimes \interp{\Gamma_{1}} \otimes \interp{V} \otimes \interp{\Gamma_{1}'} \otimes \interp{\Gamma_{2}}) \circ (\mathsf{dup}\otimes \interp{\Gamma_{1}} \otimes \interp{\Gamma_{1}'}\otimes \interp{\Gamma_{2}}\otimes \interp{\Gamma_{3}}) \]
      \[ =  \mu \circ T\epsilon  \circ \mu \circ T\sigma \circ t \circ (\interp{M} \otimes \interp{N})\]
      \[\circ (\interp{\Phi} \otimes \interp{\Gamma_{1}} \otimes \interp{V} \otimes \interp{\Gamma_{1}'} \otimes \interp{\Gamma_{2}}) \circ (\mathsf{dup}\otimes \interp{\Gamma_{1}} \otimes \interp{\Gamma_{1}'}\otimes \interp{\Gamma_{2}}\otimes \interp{\Gamma_{3}}). \]
      Thus
      \[\interp{[V/x]MN} = \interp{M N} \circ (\interp{\Phi} \otimes \interp{\Gamma_{1}} \otimes \interp{V} \otimes \interp{\Gamma_{1}'} \otimes \interp{\Gamma_{2}})\]
      \[\circ (\mathsf{dup}\otimes \interp{\Gamma_{1}} \otimes \interp{\Gamma_{1}'} \otimes \interp{\Gamma_{2}}\otimes \interp{\Gamma_{3}}).\]
    \item Case
      \begin{center}
        \begin{tabular}{l}
          \infer{\Gamma_1, x : C, \Gamma_1'
          \vdash_{\alpha \& \beta} \mathsf{force}\ M : A.}
          {\Gamma_1, x : C, \Gamma_1' \vdash_\beta M :\ !_\alpha A}
        \end{tabular}
      \end{center}
      Suppose $\Gamma_{2}\vdash_{1} V : C$ and $\alpha = \beta = 0$. By induction hypothesis, we have
      \[ \interp{[V/x]M} = \interp{M} \circ (\interp{\Gamma_{1}} \otimes \interp{V} \otimes \interp{\Gamma_{1}'})\]
      \[ : \interp{\Gamma_{1}} \otimes \interp{\Gamma_{2}} \otimes \interp{\Gamma_{1}'} \to T p\flat T \interp{A}.\]
      Since $\Gamma_{1}, \Gamma_{2}, \Gamma_{1'} \vdash_{0} \force [V/x]M : A$, we have
      \[\interp{\force [V/x]M} = \mu \circ T \force \circ \interp{[V/x]M} = \mu \circ T \force \circ \interp{M} \circ (\interp{\Gamma_{1}} \otimes \interp{V} \otimes \interp{\Gamma_{1}'})\]
      \[ : \interp{\Gamma_{1}} \otimes \interp{\Gamma_{2}} \otimes \interp{\Gamma_{1}'} \to T \interp{A}.\]
      On the other hand, we have
      \[\interp{\force M} \circ (\interp{\Gamma_{1}} \otimes \interp{V} \otimes \interp{\Gamma_{1}'})\]
      \[ = \mu \circ T \force \circ \interp{M} \circ (\interp{\Gamma_{1}} \otimes \interp{V} \otimes \interp{\Gamma_{1}'}).\]
      So $\interp{\force [V/x]M} = \interp{\force M} \circ (\interp{\Gamma_{1}} \otimes \interp{V} \otimes \interp{\Gamma_{1}'})$.
    \item Case 
      \[
        \infer{\Phi, x : P \vdash_{0} (M, N) : A\otimes B  }{\Phi, x : P \vdash_{0} M : A &
        \Phi, x : P \vdash_{0} N : B}
    \]
    Suppose $\Phi \vdash_{1} V : P$. By induction hypothesis, we have $\interp{[V/x]M} = \interp{M} \circ (\interp{\Phi}\otimes \interp{V}) \circ \mathsf{dup} : \interp{\Phi} \to T\interp{A}$ and $\interp{[V/x]N} = \interp{N} \circ (\interp{\Phi}\otimes \interp{V}) \circ \mathsf{dup} : \interp{\Phi} \to T\interp{B}$. We have
    \[
      \interp{([V/x]M, [V/x]N)} =
      \mu \circ T\sigma \circ t \circ (\interp{[V/x]M} \otimes \interp{[V/x]N})\circ \mathsf{dup}
    \]
    \[= \mu \circ T\sigma \circ t \circ ((\interp{M} \circ (\interp{\Phi}\otimes \interp{V}) \circ \mathsf{dup}) \otimes (\interp{N} \circ (\interp{\Phi}\otimes \interp{V}) \circ \mathsf{dup})\circ \mathsf{dup}.\]

    On the other hand,
    \[\interp{(M, N)} \circ (\interp{\Phi}\otimes \interp{V}) \circ \mathsf{dup}\]
    \[ = \mu \circ T\sigma \circ t \circ (\interp{M}\otimes \interp{N}) \circ \mathsf{dup}_{\interp{\Phi}\otimes \interp{P}} \circ (\interp{\Phi}\otimes \interp{V}) \circ \mathsf{dup}_{\interp{\Phi}}\]
    \[\stackrel{}{=} \mu \circ T\sigma \circ t \circ (\interp{M}\otimes \interp{N})  \circ ((\interp{\Phi}\otimes \interp{V})\otimes (\interp{\Phi}\otimes \interp{V})) \circ \mathsf{dup}_{\interp{\Phi}\otimes \interp{\Phi}} \circ \mathsf{dup}_{\interp{\Phi}}\]
    \[= \mu \circ T\sigma \circ t \circ ((\interp{M} \circ (\interp{\Phi}\otimes \interp{V}) \circ \mathsf{dup}_{\interp{\Phi}})\otimes (\interp{N} \circ (\interp{\Phi}\otimes \interp{V}) \circ \mathsf{dup}_{\interp{\Phi}}))  \circ \mathsf{dup}_{\interp{\Phi}}.\]
    Thus $\interp{([V/x]M, [V/x]N)} = \interp{(M, N)} \circ (\interp{\Phi}\otimes \interp{V}) \circ \mathsf{dup}$.
  \end{itemize}
\end{proof}

\section{Proof of Theorem \ref{thm:soundness}}
\label{app:sound}
\begin{theorem}[Soundness]
  \label{soundness}
  If $\vdash_{\alpha} (Q, M) : A; \Sigma'$, and $(Q, M) \Downarrow \sum_{i\in [n]}p_{i} (Q_{i}, V_{i})$,
    then
    \[\interp{Q, M} = \sum_{i\in [n]}p_{i} \interp{Q_{i}, V_{i}} : I \to T(\interp{A}\otimes \interp{\Sigma'}).\]
\end{theorem}
\begin{proof}
  In the case where $\alpha=1$, there is no dynamic lifting, and
  therefore the proof is mostly similar to the proof of
  Theorem~\ref{circ:sound}. We consider just one typical case:
    \[ \infer{(Q, M N)  \Downarrow (Q_{1}'', V_{1}')}
        {
        \begin{array}{cc}
          (Q, M) \Downarrow (Q_1, \lambda x. M'_1) \\
          (Q_1, N) \Downarrow (Q_{1}', V_{1})\\
          (Q_{1}', [V_{1}/x]M'_1) \Downarrow(Q_{1}'', V_{1}') 
        \end{array}
        }
      \]
      Suppose $\mathrm{addr}(Q) = \Sigma'', \Sigma'$ and
      $\Sigma'' \vdash_{1} M N : A$.
      By the typing rule for $MN$, we have
      \[
        \begin{tabular}{l}
          \infer
          {\Sigma'' \vdash_{1} M N : A,}
          {\Sigma_1'' \vdash_{1} M :  B \multimap_1 A &  \Sigma_2'' \vdash_{1} N : B}
        \end{tabular}
      \]
      where $\Sigma'' = \Sigma_{1}'', \Sigma_{2}''$.
      By Theorem \ref{thm:preserved0}, we have $\mathrm{addr}(Q_{1}) = \Sigma_{1}''', \Sigma_{2}'', \Sigma'$,
      and $\Sigma_{1}''' \vdash_{1} \lambda x. M_{1}' : B\multimap_{1} A$,
      and $\mathrm{addr}(Q_{1}') = \Sigma_{1}''', \Sigma_{2}''', \Sigma'$,
      and $\Sigma_{2}''' \vdash_{1} V_{1} : B$,
      and $\mathrm{addr}(Q_{1}'') = \Sigma'''', \Sigma'$,
      and $\Sigma''''\vdash V_{1}' : B$. 

      By induction hypothesis, we have
      \[\interp{Q, M} = T(\interp{M}\otimes \interp{\Sigma_{2}''} \otimes \interp{\Sigma'})\circ \phi Q = T(\interp{\lambda x.M_{1}'}\otimes \interp{\Sigma_{2}''} \otimes \interp{\Sigma'})\circ \phi Q_{1} \]
      \[: I \to T(\interp{B\multimap A}\otimes \interp{\Sigma_{2}''} \otimes \interp{\Sigma'})\]
      and
      \[\interp{Q_{1}, N} = T(\interp{\Sigma_{1}'''} \otimes \interp{N} \otimes \interp{\Sigma'})\circ \phi Q_{1} = T(\interp{\Sigma_{1}'''} \otimes \interp{V_{1}} \otimes \interp{\Sigma'})\circ \phi Q_{1}' \]
      \[:I \to T(\interp{\Sigma_{1}'''} \otimes \interp{B} \otimes \interp{\Sigma'})\]
      and 
      \[\interp{Q_{1}', [V_{1}/x]M_{1}'} = T(\interp{[V_{1}/x]M_{1}'}\otimes \interp{\Sigma'})\circ \phi Q_{1}' \]
      \[= T ((\interp{M_{1}'}\circ (\interp{\Sigma_{1}'''} \otimes \interp{V_{1}}))\otimes \interp{\Sigma'}) \circ \phi Q_{1}'\]
      \[= T(\interp{V_{1}'}\otimes \interp{\Sigma'})\circ \phi Q_{1}'' : I \to T(\interp{A}\otimes \interp{\Sigma'}).\]

      We need to show that
      \[\interp{Q, MN} = T(\interp{MN}\otimes \interp{\Sigma'}) \circ \phi Q\]
      \[= T((\epsilon \circ (\interp{M}\otimes \interp{N}))\otimes \interp{\Sigma'}) \circ \phi Q\]
      \[= T ((\interp{M_{1}'}\circ (\interp{V_{1}}\otimes \interp{\Sigma_{1}'''}))\otimes \interp{\Sigma'}) \circ \phi Q_{1}'.\]

      This is the case because
      \[\interp{Q, MN} = T(\interp{MN}\otimes \interp{\Sigma'}) \circ \phi Q\]
      \[= T((\epsilon \circ (\interp{M}\otimes \interp{N}))\otimes \interp{\Sigma'}) \circ \phi Q\]
      \[= T(\epsilon \otimes \interp{\Sigma'}) \circ T(\interp{B\multimap_{1} A}\otimes \interp{N}\otimes \interp{\Sigma'})\circ
        T(\interp{M}\otimes \interp{\Sigma_{2}''}\otimes \interp{\Sigma'})\circ \phi Q\]
      \[= T(\epsilon \otimes \interp{\Sigma'}) \circ T(\interp{B\multimap_{1} A}\otimes \interp{N}\otimes \interp{\Sigma'})\circ
        T(\interp{\lambda x.M_{1}'}\otimes \interp{\Sigma_{2}''}\otimes \interp{\Sigma'})\circ \phi Q_{1}\]
      \[= T(\epsilon \otimes \interp{\Sigma'}) \circ T(\interp{\lambda x.M_{1}'}\otimes \interp{N}\otimes \interp{\Sigma'}) \circ \phi Q_{1}\]
      \[ =T(\epsilon \otimes \interp{\Sigma'}) \circ T(\interp{\lambda x.M_{1}'}\otimes \interp{B}\otimes \interp{\Sigma'}) \circ
        T(\interp{\Sigma_{1}'''} \otimes \interp{N} \otimes \interp{\Sigma'}) \circ \phi Q_{1} \]
      \[= T(\epsilon \otimes \interp{\Sigma'}) \circ T(\interp{\lambda x.M_{1}'}\otimes \interp{B}\otimes \interp{\Sigma'}) \circ T(\interp{\Sigma_{1}'''}\otimes \interp{V_{1}}\otimes \interp{\Sigma'})\circ \phi Q_{1}'\]
      \[=  T(\interp{M_{1}'}\otimes \interp{\Sigma'}) \circ T(\interp{\Sigma'''_{1}} \otimes \interp{V_{1}} \otimes \interp{\Sigma'})\circ \phi Q_{1}'
      \]
      \[ = T ((\interp{M_{1}'}\circ (\interp{\Sigma_{1}'''} \otimes \interp{V_{1}}))\otimes \interp{\Sigma'}) \circ \phi Q_{1}'\]

      Now we consider the case where $\alpha=0$.  Suppose
      $\mathrm{addr}(Q) = \Sigma'', \Sigma'$ and $\Sigma'' \vdash_{0}
      M : A$. We proceed by induction on $(Q, M) \Downarrow \sum_{i\in
        [n]}p_{i} (Q_{i}, V_{i})$.

    \begin{itemize}
      
    \item Case 
      \[
        \infer{(Q, \Dyn M) \Downarrow \sum_{(i, j)\in [n] \times [2]} p_i q_{i,j} (Q_{i,j}', a_{i,j}).}
        {
          \begin{array}{ll}
            (Q, M) \Downarrow \sum_{i\in [n]} p_i (Q_i, \ell_i) \\
            \mathrm{read}(Q_i, \ell_i) = \sum_{j \in [2]}q_{i,j}(Q_{i,j}', a_{i,j})
          \end{array}
        }
      \]

      \begin{itemize}
      \item Suppose we have
        \[
          \infer{\Sigma'' \vdash_0 \Dyn M  : \Bool}
          {\Sigma'' \vdash_1 M  : \Bit}.
        \]
          In this case we have the following derivation. 
  \[\infer{(Q, \Dyn M)\Downarrow q_{1}(Q_{1}', \mathsf{False}) + q_{2}(Q_{2}', \mathsf{True})}
    {
      \begin{array}{c}
        (Q,  M) \Downarrow (Q', \ell)\\
        \mathrm{read}(Q', \ell) = q_{1}(Q_{1}', \mathsf{False}) + q_{2}(Q_{2}', \mathsf{True})
      \end{array}
    }\]

  Since $\mathrm{read}(Q', \ell) = q_{1}(Q_{1}', \mathsf{False}) +
  q_{2}(Q_{2}', \mathsf{True})$, we have
  that $Q' = q_{1}(Q_{1}' \otimes \mathrm{inj}_{1}) + q_{2}(Q_{2}' \otimes \mathrm{inj}_{2})$ in $\q$. So
  in $\A$, we have 
  \[\phi Q' = q_{1}(\mu \circ T t \circ s \circ (\phi Q_{1}'\otimes \phi(\mathrm{inj}_{1}))) + q_{2} (\mu \circ T t \circ s \circ (\phi Q_{2}'\otimes \phi(\mathrm{inj}_{2}))),\]
  where $\phi Q' : I \to T(\Bit \otimes \interp{\Sigma'})$, and
  $\phi Q_{1}', \phi Q_{2}' : I \to T \interp{\Sigma'}$, and
  $\phi(\mathrm{inj}_{1}), \phi(\mathrm{inj}_{2}) : I \to T\Bit$.
  Note that by condition \ref{embeddings}, we have $\phi(\mathrm{inj}_{1}) = \eta \circ \mathsf{init}\circ \interp{\mathsf{False}}$ and $\phi(\mathrm{inj}_{2}) = \eta \circ \mathrm{init}\circ \interp{\mathsf{True}}$. 
  We need to show that
  \[\interp{Q, \Dyn M} = q_{1}(T(\interp{\mathsf{False}}\otimes {\interp{\Sigma'}})\circ \phi Q_{1}') + q_{2}(
    T(\interp{\mathsf{True}}\otimes {\interp{\Sigma'}})\circ \phi Q_{2}').\]
  By induction hypothesis, we have $\interp{Q, M} = \interp{Q', \ell}$, i.e., $T(\interp{M}\otimes {\interp{\Sigma'}}) \circ \phi Q = \phi Q'$.
  Thus
  \[\interp{Q, \Dyn M} = \mu \circ T s \circ T((\Dyn \circ \interp{M}) \otimes {\interp{\Sigma'}})\circ \phi Q\]
  \[ = \mu \circ T s \circ T(\Dyn \otimes {\interp{\Sigma'}}  ) \circ T(\interp{M} \otimes {\interp{\Sigma'}})\circ \phi Q\]
  \[= \mu \circ T s \circ T(\Dyn \otimes {\interp{\Sigma'}}) \circ \phi Q' \]
  \[= \mu \circ T s \circ T(\Dyn \otimes {\interp{\Sigma'}})\]
  \[\circ (q_{1}(\mu \circ T t \circ s \circ (\phi Q_{1}' \otimes \phi(\mathrm{inj}_{1}) )) + q_{2} (\mu \circ T t \circ s \circ (\phi Q_{2}'\otimes \phi(\mathrm{inj}_{2}))))\]
  \[= q_{1} (\mu \circ T s \circ T(\Dyn \otimes {\interp{\Sigma'}}) \circ \mu \circ T t \circ s \circ (\phi Q_{1}'\otimes \phi(\mathrm{inj}_{1}))) \]
  \[+ q_{2} (\mu \circ T \sigma \circ T(\Dyn \otimes {\interp{\Sigma'}}) \circ \mu \circ T t \circ \sigma \circ ( \phi Q_{2}'\otimes \phi(\mathrm{inj}_{2}))).\]

  We just need to show
  \[ T(\interp{\mathrm{False}}\otimes {\interp{\Sigma'}})\circ \phi Q_{1}' = \mu \circ Ts \circ T(\Dyn \otimes {\interp{\Sigma'}}) \circ \mu \circ T t \circ s \circ (\phi Q_{1}'\otimes \phi(\mathrm{inj}_{1}))\]
  \[= \mu \circ Ts \circ T(\Dyn \otimes {\interp{\Sigma'}}) \circ \mu \circ T t \circ s \circ (\phi Q_{1}'\otimes (\eta \circ \mathsf{init}\circ \interp{\mathsf{False}})).\]
  This is true because of the following commutative diagram.

    \[ \footnotesize
      \begin{tikzcd}
        I \arrow[d, "\phi Q_{1}'"]&&&& \\
        T\interp{\Sigma'} \arrow[d, "\lambda^{-1}"] \arrow[dr, "T\lambda^{-1}"] & & & & \\
        I \otimes T\interp{\Sigma'} \arrow[r, "t"]
        \arrow[d, "\interp{\mathsf{False}} \otimes T\interp{\Sigma'}"]& T(I \otimes \interp{\Sigma'})
        \arrow[d, "T(\interp{\mathsf{False}} \otimes \interp{\Sigma'})"]&&& \\
        \Bool \otimes T\interp{\Sigma'}  \arrow[r, "t"]
        \arrow[d, "\mathrm{init} \otimes T\interp{\Sigma'}"]
        & T( \Bool \otimes \interp{\Sigma'})
        \arrow[rr, "\id", bend left = 15]
        \arrow[d, "T(\mathrm{init} \otimes \interp{\Sigma'})"]
        \arrow[dr, "T(\eta \otimes \interp{\Sigma'})", bend left = 5]
        \arrow[drr, "\eta", bend left = 15]
        && T(\Bool \otimes \interp{\Sigma'} ) & \\
        \Bit \otimes T\interp{\Sigma'} 
        \arrow[r, "t"]
        \arrow[d, "\eta \otimes T\interp{\Sigma'}", swap]
        & T(\Bit \otimes \interp{\Sigma'} )
        \arrow[ddl, "T(\eta \otimes \interp{\Sigma'})", near start, swap]
        \arrow[dddl, "\eta", bend left = 10]
        \arrow[r, "T( \Dyn\otimes \interp{\Sigma'})", swap, bend right = 10]
        &T(T\Bool \otimes \interp{\Sigma'} ) \arrow[r, "T s"]
        & TT(\Bool\otimes \interp{\Sigma'}) \arrow[u, "\mu"]& \\
        T \Bit \otimes T\interp{\Sigma'}  \arrow[d, "t"]& & & & \\
        T(T \Bit \otimes \interp{\Sigma'}) \arrow[d, "Ts"] &&&& \\
        TT(\Bit \otimes \interp{\Sigma'} ) \arrow[r, "\mu"]& T(  \Bit \otimes \interp{\Sigma'}) \arrow[uuu, "\id"]&&&
      \end{tikzcd}
    \]

      \item Suppose we have
        \[
          \infer{\Sigma'' \vdash_0 \Dyn M  : \Bool}
          {\Sigma'' \vdash_0 M  : \Bit}.
        \]
        In this case $(Q, M) \Downarrow \sum_{i\in [n]}p_{i}(Q_{i}, \ell_{i})$ and $\mathrm{read}(Q_{i}, \ell_{i}) = q_{i,1}(Q_{i,1}', \mathrm{False}) + q_{i,2}(Q_{i,2}', \mathrm{True})$, where $\addr{Q_{i}} = \Sigma', \ell_{i} : \Bit$ and $\addr{Q_{i,1}'} = \addr{Q_{i,1}'} = \Sigma'$.
        Thus we have the following in $\A$:
        $\phi Q_{i} = q_{i,1} (\mu \circ T t \circ s \circ (\phi Q_{i, 1}'\otimes \phi(\mathrm{inj}_{1}))) + q_{i,2}(\mu \circ T t \circ s \circ (\phi Q_{i, 2}' \otimes \phi(\mathrm{inj_{2}})))$. Note that $\phi(\mathrm{inj}_{1}) = \eta \circ \mathsf{init}\circ \interp{\mathsf{False}}$ and $\phi(\mathrm{inj}_{2}) = \eta \circ \mathrm{init}\circ \interp{\mathsf{True}}$. 

        We need to show that \[\interp{Q, \Dyn M}\]
        \[ = \sum_{i\in [n]}p_{i}q_{i,1}(T(\interp{\mathrm{False}}\otimes \interp{\Sigma'})\otimes \phi Q_{i,1}') + p_{i}q_{i,2}(T(\interp{\mathrm{True}}\otimes \interp{\Sigma'})\otimes \phi Q_{i,2}').\]
        By induction hypothesis, we have
        \[ \interp{Q, M} = \mu \circ T s \circ T (\interp{M} \otimes {\interp{\Sigma'}})\circ \phi Q\]
        \[ = \sum_{i \in [n]}p_{i}(T(\interp{\ell_{i}}\otimes {\interp{\Sigma'}})\circ \phi Q_{i}) = \sum_{i \in [n]}p_{i} \phi Q_{i}.\]
        Thus
        \[ \interp{Q, \Dyn M} = \mu \circ T s \circ T((\mu \circ T\Dyn \circ \interp{M}) \otimes {\interp{\Sigma'}})\circ \phi Q \]
        \[= \mu \circ T s \circ T((\mu \circ T\Dyn) \otimes {\interp{\Sigma'}})  \circ T( \interp{M} \otimes {\interp{\Sigma'}})\circ \phi Q\]
        \[\stackrel{(*)}{=} \mu \circ T s \circ T(\Dyn \otimes {\interp{\Sigma'}})  \circ \mu \circ T s \circ T( \interp{M} \otimes {\interp{\Sigma'}})\circ \phi Q\]
        \[= \mu \circ T s \circ T(\Dyn \otimes {\interp{\Sigma'}})  \circ (\sum_{i \in [n]}p_{i} \phi Q_{i})\]
        \[ = \sum_{i \in [n]}p_{i} (\mu \circ T s \circ T(\Dyn \otimes {\interp{\Sigma'}})  \circ \phi Q_{i})\]
        \[=  \sum_{i \in [n]}p_{i} (\mu \circ T s \circ T(\Dyn \otimes {\interp{\Sigma'}})  \circ (q_{i,1} (\mu \circ T t \circ s \circ (\phi Q_{i, 1}'\otimes \phi(\mathrm{inj_{1}})))\]
        \[  + q_{i,2}(\mu \circ T t \circ s \circ (\phi Q_{i, 2}'\otimes \phi(\mathrm{inj_{2}})))))\]
        \[ = \sum_{i \in [n]} p_{i} q_{i,1}  (\mu \circ T s \circ T(\Dyn \otimes {\interp{\Sigma'}})  \circ \mu \circ T t \circ s \circ (\phi Q_{i, 1}'\otimes \phi(\mathrm{inj_{1}})))\]
        \[ + p_{i} q_{i,2} (\mu \circ T s \circ T(\Dyn \otimes {\interp{\Sigma'}})  \circ\mu \circ T t \circ s \circ (\phi Q_{i, 2}'\otimes
          \phi(\mathrm{inj_{2}}))).\]

        So we just need to show
        \[ T(\interp{\mathrm{False}}\otimes \interp{\Sigma'})\otimes \phi Q_{i,1}'\]
        \[ = \mu \circ T s \circ T(\Dyn \otimes {\interp{\Sigma'}})  \circ \mu \circ T t \circ s \circ (\phi Q_{i, 1}'\otimes \phi(\mathrm{inj_{1}})).\]
        We already showed this in the previous case.

        The equality of $(*)$ is by the following commutative diagram.

        \[ \footnotesize
          \begin{tikzcd}[column sep=1.8em]
            && T(T\Bit \otimes \interp{\Sigma'})
            \arrow[dr, "T( T\Dyn \otimes \interp{\Sigma'})"]
            \arrow[dl, "Ts"]
            && \\
            &TT( \Bit \otimes \interp{\Sigma'})
            \arrow[dl, "\mu"]
            \arrow[dr, "TT( \Dyn \otimes \interp{\Sigma'})"]&
            & T(TT\Bool \otimes \interp{\Sigma'} )
            \arrow[dr, "T(\mu \otimes \interp{\Sigma'})"]
            \arrow[dl, "Ts"]
            & \\
            T(\Bit \otimes \interp{\Sigma'})
            \arrow[dr, "T( \Dyn\otimes \interp{\Sigma'})"]&&
            TT(T\Bool \otimes \interp{\Sigma'})
            \arrow[dl, "\mu"] \arrow[d, "TTs"]
            && T(T\Bool \otimes \interp{\Sigma'})
            \arrow[d, "Ts"]\\
            & T( T\Bool \otimes \interp{\Sigma'})
            \arrow[d, "Ts"]
            & TTT(\Bool \otimes \interp{\Sigma'})
            \arrow[rr, "\mu"]
            \arrow[dl, "\mu"]
            && TT(\Bool \otimes \interp{\Sigma'})
            \arrow[dl, "\mu"]
            \\
            & TT(\Bool\otimes \interp{\Sigma'} )
            \arrow[rr, "\mu"]
            &&T(\Bool \otimes \interp{\Sigma'}) &
          \end{tikzcd}
        \]

    \end{itemize}

  \item Case

    \[
      \infer{(Q, M N)  \Downarrow \sum_{(i,j,k)\in [n]\times [m]\times [l]} p_i q_{i,j} s_{i,j,k} (Q_{i,j,k}'', V_{i,j,k}').}
      {
        \begin{array}{cc}
          (Q, M) \Downarrow \sum_{i\in [n]} p_i(Q_i, \lambda x. M'_i) \\
          (Q_i, N) \Downarrow \sum_{j\in [m]} q_{i,j} (Q_{i,j}', V_{i,j})\\
          (Q_{i,j}', [V_{i,j}/x]M'_i) \Downarrow \sum_{k\in [l]} s_{i,j, k} (Q_{i,j,k}'', V_{i,j,k}') 
        \end{array}
      }
    \]

    Here we only consider the following typing judgment (the other cases are proved similarly). 
    \[
      \begin{tabular}{lll}
        \infer
        {\Sigma_1, \Sigma_2 \vdash_{0} M N : B}
        {\Sigma_1 \vdash_{0} M :  A \multimap_0 B & \Sigma_2 \vdash_{0} N : A}
      \end{tabular}
    \]
    We assume that $\addr{Q} = \Sigma_{1}, \Sigma_{2}, \Sigma'$.

    By induction hypothesis, we have
    \[ \interp{Q, M} = \mu \circ Ts \circ T(s \otimes \interp{\Sigma'}) \circ T( \interp{M} \otimes \interp{\Sigma_2}\otimes \interp{\Sigma'}) \circ \phi Q\]
    \[ = \sum_{i\in [n]} p_{i}\interp{Q_{i}, \lambda x.M_{i}'}
      = \sum_{i\in [n]} p_{i} (T(\interp{\lambda x. M'_{i}} \otimes \interp{\Sigma_2}\otimes \interp{\Sigma'})\circ \phi Q_i),\]

    where $\phi Q_i : I \to T(\interp{\Sigma_i'}\otimes \interp{\Sigma_2}\otimes \interp{\Sigma'})$ and $\Sigma_i' \vdash_1 \lambda x. M'_{i} : A \multimap_0 B$. 
    The induction hypothesis also gives
    \[ \interp{Q_i, N} = \mu \circ Ts \circ T (t\otimes \interp{\Sigma'}) \circ T (\interp{\Sigma_i'}\otimes  \interp{N} \otimes \interp{\Sigma'})\circ \phi Q_i\]
    \[= \sum_{j\in [m]} q_{i, j} \interp{Q'_{i,j}, V_{i,j}}
      = \sum_{j\in [m]} q_{i, j} (T(\interp{\Sigma_i'} \otimes \interp{V_{i,j}}\otimes \interp{\Sigma'})\circ \phi Q_{i,j}'),\]
    where $Q_{i,j}' : I \to T(\interp{\Sigma_i'} \otimes \interp{\Sigma_{i,j}''}\otimes \interp{\Sigma'})$ and
    $\Sigma_{i,j}'' \vdash_1 V_{i,j} : A$.
    Moreover,
    \[\interp{Q_{i,j}', [V_{i,j}/x]M_{i}'} = \mu \circ Ts \circ T (\interp{[V_{i,j}/x]M_{i}'} \otimes \interp{\Sigma'})\circ \phi Q_{i,j}'
    \]
    \[= \mu \circ Ts \circ T((\interp{M_{i}'} \circ ( \interp{\Sigma_{i}'} \otimes \interp{V_{i,j}} )) \otimes \interp{\Sigma'})\circ \phi Q_{i,j}'\]
    \[= \sum_{k\in [l]} s_{i,j,k}\interp{Q_{i,j,k}'', V'_{i,j,k}}
      = \sum_{k\in [l]} s_{i,j,k} (T(\interp{V'_{i,j,k}} \otimes \interp{\Sigma'})\circ \phi Q_{i,j,k}''),\]
    where $\phi Q_{i,j,k}'' : I \to T(\interp{\Sigma'''_{i,j,k}} \otimes \interp{\Sigma'})$ and $\Sigma'''_{i,j,k} \vdash_1 V'_{i,j,k} : B$.

    We need to show
    \[ \interp{Q, MN} = \mu \circ Ts \circ T(\interp{M N} \otimes \interp{\Sigma'})\circ \phi Q\]
    \[= \mu \circ Ts \circ T((\mu \circ T\epsilon \circ \mu \circ Ts \circ t \circ (\interp{M}\otimes \interp{N})) \otimes \interp{\Sigma'})
      \circ \phi Q\]
    \[= \sum_{i,j,k\in [n]\times [m]\times [l]} p_{i}q_{i,j}s_{i,j,k}(T(\interp{V'_{i,j,k}} \otimes \interp{\Sigma'})\circ \phi Q''_{i,j,k}).\]
    On the right hand side, we have
    \[RHS = \sum_{i,j,k\in [n]\times [m]\times [l]}  p_{i}q_{i,j}s_{i,j,k}(T(\interp{V'_{i,j,k}} \otimes \interp{\Sigma'})\circ \phi Q''_{i,j,k}) \]
    \[= \sum_{i,j\in [n]\times [m]}  p_{i}q_{i,j} \sum_{k\in [l]} s_{i,j,k}(T(\interp{V'_{i,j,k}} \otimes \interp{\Sigma'})\circ \phi Q''_{i,j,k})\]
    \[ \stackrel{IH}{=} \sum_{i,j\in [n]\times [m]}  p_{i}q_{i,j}  (\mu \circ Ts \circ T((\interp{M_{i}'} \circ ( \interp{\Sigma_{i}'} \otimes \interp{V_{i,j}} )) \otimes \interp{\Sigma'})\circ \phi Q_{i,j}'). \]

    On the left hand side, we have

    \[LHS = \mu \circ Ts \circ T((\mu \circ T\epsilon \circ \mu \circ Ts \circ t \circ (\interp{M}\otimes \interp{N})) \otimes \interp{\Sigma'})\circ \phi Q\]
    \[ =  \mu \circ Ts \circ T((\mu \circ T\epsilon \circ \mu \circ T s \circ t \circ (T(\interp{A}\multimap T\interp{B})\otimes \interp{N}) \circ (\interp{M}\otimes \interp{\Sigma_{2}})) \otimes \interp{\Sigma'})\circ \phi Q\]
    \[ =  \mu \circ Ts \circ T(\mu \circ T\epsilon \circ \mu \circ Ts \circ t \circ (T(\interp{A}\multimap T\interp{B})\otimes \interp{N}) \otimes \interp{\Sigma'}) \circ T(\interp{M}\otimes \interp{\Sigma_{2}} \otimes \interp{\Sigma'}) \circ \phi Q \]
    \[\stackrel{(1)}{=} \mu \circ Ts \circ T(\mu \otimes \interp{\Sigma'}) \circ T(T\epsilon \otimes \interp{\Sigma'})
      \circ T(t \otimes \interp{\Sigma'}) \circ T(\interp{A} \multimap T\interp{B} \otimes \interp{N} \otimes \interp{\Sigma'})
    \]
    \[\circ \mu \circ Ts \circ T(s \otimes \interp{\Sigma'}) \circ T(\interp{M}\otimes \interp{\Sigma_{2}} \otimes \interp{\Sigma'}) \circ \phi Q \]
    \[\stackrel{IH}{=} \mu \circ Ts \circ T(\mu \otimes \interp{\Sigma'}) \circ T(T\epsilon \otimes \interp{\Sigma'})
      \circ T(t \otimes \interp{\Sigma'}) \circ T(\interp{A} \multimap T\interp{B} \otimes \interp{N} \otimes \interp{\Sigma'}) \circ\]
    \[\sum_{i\in [n]} p_{i} (T(\interp{\lambda x. M'_{i}} \otimes \interp{\Sigma_2}\otimes \interp{\Sigma'})\circ \phi Q_i)\]
    \[= \sum_{i\in [n]} p_{i} (\mu \circ Ts \circ T(\mu \otimes \interp{\Sigma'}) \circ T(T\epsilon \otimes \interp{\Sigma'})
      \circ T(t \otimes \interp{\Sigma'}) \circ T(\interp{A} \multimap T\interp{B} \otimes \interp{N} \otimes \interp{\Sigma'}) \circ\]
    \[ T(\interp{\lambda x. M'_{i}} \otimes \interp{\Sigma_2}\otimes \interp{\Sigma'})\circ \phi Q_i)\]
    \[ = \sum_{i\in [n]} p_{i} (\mu \circ Ts \circ T((\mu \circ T\epsilon \circ t \circ ((\interp{A}\multimap T\interp{B}) \otimes \interp{N}) \circ (\interp{\lambda x.M_{i}'} \otimes \interp{\Sigma_{2}}))\otimes \interp{\Sigma'}) \circ \phi Q_{i})\]
    \[ = \sum_{i\in [n]} p_{i} (\mu \circ Ts \circ T((\mu \circ T\epsilon \circ t \circ (\interp{\lambda x.M_{i}'} \otimes \interp{N}))\otimes \interp{\Sigma'}) \circ \phi Q_{i})\]
    \[= \sum_{i\in [n]} p_{i} (\mu \circ Ts \circ T((\mu \circ T\epsilon \circ t \circ (\interp{\lambda x.M_{i}'} \otimes T\interp{A}) \circ (\interp{\Sigma_{i}'} \otimes \interp{N}))\otimes \interp{\Sigma'}) \circ \phi Q_{i}) \]
    \vspace{-0.75em}
    \[= \sum_{i\in [n]} p_{i} (\mu \circ Ts \circ T((\mu \circ T\epsilon \circ t \circ (\interp{\lambda x.M_{i}'} \otimes T\interp{A}))\otimes \interp{\Sigma'}) \circ T(\interp{\Sigma_{i}'} \otimes \interp{N} \otimes \interp{\Sigma'}) \circ \phi Q_{i}) \]
    \vspace{-0.75em}
    \[\stackrel{(2)}{=} \sum_{i\in [n]} p_{i} (\mu \circ Ts \circ T(\epsilon \otimes \interp{\Sigma'}) \circ T(\interp{\lambda x.M_{i}'} \otimes \interp{A}\otimes \interp{\Sigma'}) \circ \mu \circ Ts \circ T(t \otimes \interp{\Sigma'}) \circ T(\interp{\Sigma_{i}'} \otimes \interp{N} \otimes \interp{\Sigma'}) \circ \phi Q_{i})\]
    \vspace{-0.75em}
    \[\stackrel{IH}{=} \sum_{i\in [n]} p_{i} (\mu \circ Ts \circ T(\epsilon \otimes \interp{\Sigma'}) \circ T(\interp{\lambda x.M_{i}'} \otimes \interp{A}\otimes \interp{\Sigma'}) \circ \sum_{j\in [m]} q_{i, j} (T(\interp{\Sigma_i'} \otimes \interp{V_{i,j}}\otimes \interp{\Sigma'})\circ \phi Q_{i,j}'))\]
    \vspace{-0.75em}
    \[= \sum_{i\in [n]} p_{i} (\mu \circ Ts \circ  T(\interp{M_{i}'} \otimes \interp{\Sigma'}) \circ \sum_{j\in [m]} q_{i, j} (T(\interp{\Sigma_i'} \otimes \interp{V_{i,j}}\otimes \interp{\Sigma'})\circ \phi Q_{i,j}'))\]
    \vspace{-0.75em}
    \[= \sum_{i,j\in [n]\times [m]}  p_{i}q_{i,j}  (\mu \circ Ts \circ T((\interp{M_{i}'} \circ ( \interp{\Sigma_{i}'} \otimes \interp{V_{i,j}} )) \otimes \interp{\Sigma'})\circ \phi Q_{i,j}'). \]

    To show $(1)$ holds, we just need to  
    show \[ T((\mu \circ T\epsilon \circ \mu \circ Ts \circ t \circ (T(\interp{A}\multimap T\interp{B})\otimes \interp{N}) \otimes \interp{\Sigma'}) \]
    \[=  T(\mu \otimes \interp{\Sigma'}) \circ T(T\epsilon \otimes \interp{\Sigma'})
      \circ T(t \otimes \interp{\Sigma'}) \circ T(\interp{A} \multimap T\interp{B} \otimes \interp{N} \otimes \interp{\Sigma'})
      \circ \mu \circ Ts \circ T(s \otimes \interp{\Sigma'}).\]
    This is true because the following diagram commutes.

    \[

\begin{tikzcd}[column sep=-6em,row sep=3em]
  &&& T((T(\interp{A}\multimap T\interp{B})\otimes \interp{\Sigma_{2}})\otimes \interp{\Sigma'}) \arrow[dl, "T(s \otimes \interp{\Sigma'})"] \arrow[dr, "T((T(\interp{A}\multimap T\interp{B})\otimes \interp{N})\otimes \interp{\Sigma'})"]&& \\
  &&T(T((\interp{A}\multimap T\interp{B})\otimes \interp{\Sigma_{2}})\otimes \interp{\Sigma'})
  \arrow[dl, "Ts"]
  \arrow[dr, "T(T((\interp{A}\multimap T\interp{B})\otimes \interp{N})\otimes \interp{\Sigma'})"]& & T((T(\interp{A}\multimap T\interp{B})\otimes T\interp{A})\otimes \interp{\Sigma'})
  \arrow[dl, "T(s \otimes \interp{\Sigma'})"]
  \arrow[dr, "T(t \otimes \interp{\Sigma'})"]& \\
  & TT(((\interp{A}\multimap T\interp{B})\otimes \interp{\Sigma_{2}})\otimes \interp{\Sigma'})
  \arrow[dr, "TT(((\interp{A}\multimap T\interp{B})\otimes \interp{N})\otimes \interp{\Sigma'})"]
  \arrow[dl, "\mu"]&&T(T((\interp{A}\multimap T\interp{B})\otimes T\interp{A})\otimes \interp{\Sigma'})
  \arrow[dr, "T(T t \otimes \interp{\Sigma'})"]
  \arrow[dl, "Ts"]&& T(T(T(\interp{A}\multimap T\interp{B})\otimes \interp{A})\otimes \interp{\Sigma'})
  \arrow[dl, "T(Ts \otimes \interp{\Sigma'})"]\\
  T(((\interp{A}\multimap T\interp{B})\otimes \interp{\Sigma_{2}})\otimes \interp{\Sigma'})
  \arrow[dr, "T(((\interp{A}\multimap T\interp{B})\otimes \interp{N})\otimes \interp{\Sigma'})"] && TT(((\interp{A}\multimap T\interp{B})\otimes T\interp{A})\otimes \interp{\Sigma'})
  \arrow[dl, "\mu"]
  \arrow[dr, "TT(t\otimes \interp{\Sigma'})"]&& T(TT((\interp{A}\multimap T\interp{B})\otimes \interp{A})\otimes \interp{\Sigma'})
  \arrow[dl, "Ts"]
  \arrow[dr, "T(\mu \otimes \interp{\Sigma'})"]& \\
  &T(((\interp{A}\multimap T\interp{B})\otimes T\interp{A})\otimes \interp{\Sigma'})
  \arrow[rrrr, bend right = 10, "T(t\otimes \interp{\Sigma'})"]&&TT(T((\interp{A}\multimap T\interp{B})\otimes \interp{A})\otimes \interp{\Sigma'})
  \arrow[rr, "\mu"]& & T(T((\interp{A}\multimap T\interp{B})\otimes \interp{A})\otimes \interp{\Sigma'})
\end{tikzcd}

    \]

    To show $(2)$ holds, we need to show
    \[ T((\mu \circ T\epsilon \circ t \circ (\interp{\lambda x.M_{i}'} \otimes T\interp{A}))\otimes \interp{\Sigma'})\]
    \[=  T(\epsilon \otimes \interp{\Sigma'}) \circ T(\interp{\lambda x.M_{i}'} \otimes \interp{A}\otimes \interp{\Sigma'}) \circ \mu \circ Ts \circ T(t \otimes \interp{\Sigma'}).\]
 
    This is true because the following diagram commutes.
      
    \[

\begin{tikzcd}[column sep=-6em, row sep=1.5em]
  &&
  &T(\interp{\Sigma_i'} \otimes T\interp{A} \otimes \interp{\Sigma'})
  \arrow[dr, "T(\interp{\lambda x.M_i} \otimes T\interp{A} \otimes \interp{\Sigma'})"]
  \arrow[dl, "T(t\otimes \interp{\Sigma'})"]& \\
  &
  & T(T(\interp{\Sigma_i'} \otimes \interp{A}) \otimes \interp{\Sigma'})
  \arrow[dl, "Ts"]
  \arrow[dr, "T(T(\interp{\lambda x.M_i} \otimes \interp{A}) \otimes \interp{\Sigma'})"]& & T((\interp{A} \multimap T\interp{B}) \otimes T\interp{A} \otimes \interp{\Sigma'})
  \arrow[dl, "T(t\otimes \interp{\Sigma'})"]\\
  & TT((\interp{\Sigma_i'} \otimes \interp{A}) \otimes \interp{\Sigma'})
  \arrow[dr, "TT((\interp{\lambda x.M_i'} \otimes \interp{A}) \otimes \interp{\Sigma'})"]
  \arrow[dl, "\mu"]&& T(T((\interp{A}\multimap T \interp{B}) \otimes \interp{A}) \otimes \interp{\Sigma'})
  \arrow[dl, "Ts"]
  \arrow[dr, "T(T\epsilon \otimes \interp{\Sigma'})"]&  \\
  T((\interp{\Sigma_i'} \otimes \interp{A}) \otimes \interp{\Sigma'})
  \arrow[dr, "T((\interp{\lambda x.M_i'} \otimes \interp{A}) \otimes \interp{\Sigma'})"]&  & TT(((\interp{A}\multimap T \interp{B}) \otimes \interp{A}) \otimes \interp{\Sigma'})
  \arrow[dl, "\mu"]
  \arrow[dr, "TT(\epsilon \otimes \interp{\Sigma'})"]&
  & T(TT\interp{B}\otimes \interp{\Sigma'})
  \arrow[dl, "Ts"]
  \arrow[ddll, bend left = 20, "T(\mu \otimes \interp{\Sigma'})"]\\
  & T(((\interp{A} \multimap T\interp{B}) \otimes \interp{A}) \otimes \interp{\Sigma'}) \arrow[dr, "T(\epsilon\otimes \interp{\Sigma'})"]&& TT(T \interp{B} \otimes \interp{\Sigma'})
              \arrow[dl, "\mu"]& \\
  && T(T \interp{B} \otimes \interp{\Sigma'}) &&
\end{tikzcd}

    \]

  \item Case
    \[
      \infer{(Q, \mathsf{box}\ S\ M) \Downarrow \sum_{i\in [n]}p_i(Q_i, (a,\dD_i,b_i)). }
      {
        \begin{array}{cc}
          (Q, M) \Downarrow \sum_{i\in [n]}p_i(Q_i, \mathsf{lift} \ M_i') \\
          \mathrm{gen}(S) = a \\
          (\id_{\Sigma_{a}}, M_i'\ a) \Downarrow (\dD_i, b_i)
        \end{array}
      }
    \]

    Suppose $\addr{Q} = \Sigma_{1}, \Sigma'$. Consider the following.
    \[
      \infer{\Sigma_{1} \vdash_0 \mathsf{box}\ S \  M : \mathbf{Circ}(S,U)}{\Sigma_{1} \vdash_0 M : {!}_1 (S \multimap_1 U)}
    \]
    By induction hypothesis, we have
    \[\interp{Q, M} = \mu \circ Ts \circ T(\interp{M}\otimes \interp{\Sigma'})\circ \phi Q = \sum_{i\in [n]}p_{i} \interp{Q_{i}, \mathsf{lift}M_{i}'}\]
    \[= \sum_{i\in [n]}p_{i}   T(\interp{\mathsf{lift}M'_{i}}\otimes \interp{\Sigma'})  \circ \phi Q_{i}\]

    where $\phi Q_{i} : I \to T \interp{\Sigma'}$
    and $\empty \vdash_{1} \lift M_{i}' : {!}_{1}(S \multimap_{1} U)$.
    Note that $\Sigma_{a} \vdash a : S$. By Theorem \ref{circ:sound}, we have 
    \[ \interp{\id_{\Sigma_{a}}, M_{i}' a} = \epsilon \circ (\interp{M_{i}'}\otimes \interp{a}) = \interp{b_{i}}\circ \dD_{i} : \interp{\Sigma_{a}} \to \interp{U}.\]
    We need to show
    \[\interp{Q, \mathsf{box}_{S} M} = \mu \circ Ts \circ T(\interp{\mathsf{box}_{S} M} \otimes \interp{\Sigma'})\circ \phi Q\]
    \[= \mu \circ Ts \circ T((T\mathsf{box}\circ \interp{M}) \otimes \interp{\Sigma'})\circ \phi Q\]
    \[= \sum_{i\in[n]}p_{n}\interp{Q_{i}, (a, D_{i}, b)}\]
    \[ =
      \sum_{i\in [n]}p_{n} (T((\mathsf{box}\circ p \delta{\curry{\theta_{\dD_{i}}}}) \otimes \interp{\Sigma'}) \circ \phi Q_{i}),\]

    where $\theta_{\dD_{i}} = \interp{b_{i}}\circ \dD_{i} \circ \interp{a}^{-1} : \interp{S} \to \interp{U}$, and $\curry{\theta_{\dD_{i}}} : I \to \interp{S}\multimap \interp{U}$ and $\delta{\curry{\theta_{\dD_{i}}}} : 1 \to \flat (\interp{S}\multimap \interp{U})$.

  This is true because we have the following.
  \[\mu \circ Ts \circ T((T\mathsf{box}\circ \interp{M}) \otimes \interp{\Sigma'})\circ \phi Q\]
  \[ = T(\mathsf{box}\otimes \interp{\Sigma'}) \circ \mu \circ Ts \circ T(\interp{M} \otimes \interp{\Sigma'}) \circ \phi Q \]
  \[\stackrel{IH}{=} T(\mathsf{box}\otimes \interp{\Sigma'}) \circ \sum_{i\in [n]}p_{i}   (T(\interp{\mathsf{lift}M'_{i}}\otimes \interp{\Sigma'})  \circ \phi Q_{i})\]
  \[= \sum_{i\in [n]}p_{i} (T(\mathsf{box}\otimes \interp{\Sigma'}) \circ T(\interp{\mathsf{lift}M'_{i}}\otimes \interp{\Sigma'})  \circ \phi Q_{i})
  \]
    \[= \sum_{i\in [n]}p_{n} (T((\mathsf{box}\circ p\delta\interp{M_{i}'}) \otimes \interp{\Sigma'}) \circ \phi Q_{i})\]
  \[= \sum_{i\in [n]}p_{n} (T((\mathsf{box}\circ p\delta\curry{\theta_{\dD_{i}}}) \otimes \interp{\Sigma'}) \circ \phi Q_{i}).\]
  The third equality holds because $T(f)\circ g$ is linear in $g$;
  this is a consequence of the bilinearity of composition in the
  Kleisli category.
  The last equality holds because
  \[\curry{\theta_{\dD_{i}}} = \curry{\interp{b_{i}}\circ \dD_{i} \circ \interp{a}^{-1}}\]
  \[ = \curry{(\epsilon \circ (\interp{M_{i}'}\otimes \interp{a})) \circ \interp{a}^{-1}} = \curry{\epsilon \circ (\interp{M_{i}'}\otimes \interp{S})} = \interp{M_{i}'}.\]

\item Case
  \[
    \infer{(Q, \force M)  \Downarrow \sum_{(i,j)\in [n]\times [m]}p_i q_{i,j} (Q_{i,j}', V_{i,j}).}
    {
      \begin{array}{ll}
        (Q, M) \Downarrow \sum_{i\in [n]}p_i (Q_i, \lift M_i')
        \\ (Q_i, M_i') \Downarrow \sum_{j\in [m]}q_{i,j}(Q_{i,j}', V_{i,j}) 
      \end{array}}
  \]

  Suppose $\addr{Q} = \Sigma_{1}, \Sigma'$
  and 
  \[
    \infer{\Sigma_{1} \vdash_{0} \mathsf{force}\ M : A}
    {\Sigma_{1} \vdash_0 M :\ !_0 A}.
  \]

  By induction hypothesis, we have
  \[\interp{Q, M} = \mu \circ Ts \circ (\interp{M}\otimes \interp{\Sigma'})\circ \phi Q = \sum_{i\in [n]} p_{i}\interp{Q_{i}, \lift M_{i}'}\]
  \[= \sum_{i\in [n]} p_{i} (T(\interp{\lift M_{i}'}\otimes \interp{\Sigma'}) \circ \phi Q_{i}),\]
  where $\phi Q_{i} : I \to T\interp{\Sigma'}$ and $\emptyset \vdash_{1} \lift M_{i} : {!}_{0}A$.

  Also by induction hypothesis, we have
  \[ \interp{Q_{i}, M_{i}'} = \mu \circ Ts \circ (\interp{M_{i}'}\otimes \interp{\Sigma'})\circ \phi Q_{i} = \sum_{j\in[m]}q_{i,j}\interp{Q_{i,j}', V_{i,j}}\]
  \[= \sum_{j\in[m]}q_{i,j} (T(\interp{V_{i,j}}\otimes \interp{\Sigma'}) \circ \phi Q_{i,j}'),\]
  where $\phi Q_{i,j}' : I \to T(\interp{\Sigma_{2}}\otimes \interp{\Sigma'})$ and $\Sigma_{2} \vdash_{1} V_{i,j} : A$. 

  We need to show
  \[\interp{Q, \force M} = \mu \circ Ts \circ T((\mu \circ T\force \circ \interp{M})\otimes \interp{\Sigma'})\circ \phi Q\]
  \[= \sum_{i,j\in [n]\times [m]}p_{i}q_{i,j}\interp{Q_{i,j}', V_{i,j}} = \sum_{i,j\in [n]\times [m]}p_{i}q_{i,j} T(\interp{V_{i,j}}\otimes \interp{\Sigma'}) \circ \phi Q_{i,j}'.\]

  This is true because
  \[ \mu \circ Ts \circ T((\mu \circ T\force \circ \interp{M})\otimes \interp{\Sigma'})\circ \phi Q \]
  \[ = \mu \circ Ts \circ T((\mu \circ T\force)\otimes \interp{\Sigma'}) \circ T(\interp{M}\otimes \interp{\Sigma'}) \circ \phi Q\]
  \[=  \mu \circ Ts \circ T(\force\otimes \interp{\Sigma'})\circ \mu \circ Ts \circ T(\interp{M}\otimes \interp{\Sigma'}) \circ \phi Q\]
  \[ \stackrel{IH}{=} \mu \circ Ts \circ T(\force\otimes \interp{\Sigma'})\circ \sum_{i\in [n]} p_{i} (T(\interp{\lift M_{i}'}\otimes \interp{\Sigma'}) \circ \phi Q_{i})\]
  \[= \sum_{i\in [n]} p_{i} (\mu \circ Ts \circ T(\force\otimes \interp{\Sigma'})\circ T(\interp{\lift M_{i}'}\otimes \interp{\Sigma'}) \circ \phi Q_{i})\]
  \[= \sum_{i\in [n]} p_{i} (\mu \circ Ts \circ T(\interp{M_{i}'}\otimes \interp{\Sigma'})\circ \phi Q_{i})\]
  \[ \stackrel{IH}{=} \sum_{i\in [n]} p_{i} \sum_{j\in[m]}q_{i,j} (T(\interp{V_{i,j}}\otimes \interp{\Sigma'}) \circ \phi Q_{i,j}')\]
  \[ = \sum_{i,j\in [n]\times [m]} p_{i} q_{i,j} (T(\interp{V_{i,j}}\otimes \interp{\Sigma'}) \circ \phi Q_{i,j}').\]

\item Case
  \[
    \infer{(Q, \mathsf{apply}(M, N))  \Downarrow \sum_{(i,j)\in [n]\times [m]}p_i q_{i,j} (Q_{i,j}'', b_i).}
    {
      \begin{array}{cc}
        (Q, M) \Downarrow \sum_{i\in [n]}p_i (Q_i, (a_{i}, \dD_i, b_i)) \\
        (Q_i, N) \Downarrow \sum_{j\in [m]}q_{i,j}(Q_{i,j}', V_{i,j})\\
        \mathrm{operate}(\dD_i,  Q_{i,j}', a_{i}, V_{i,j}) = Q_{i,j}''
      \end{array}
    }
  \]

  Suppose $\addr{Q} = \Sigma_{1}, \Sigma_{2}, \Sigma'$ and 
  \[
    \infer{\Sigma_1, \Sigma_2 \vdash_{0} \mathsf{apply}(M, N) :  U.}{\Sigma_1 \vdash_0 M : \mathbf{Circ}(S,U) & \Sigma_2 \vdash_0 N : S}
  \]

  By induction hypothesis, we have
  \[ \interp{Q, M} = \mu \circ Ts \circ T(s \otimes \interp{\Sigma'}) \circ T(\interp{M} \otimes \interp{\Sigma_{2}}\otimes \interp{\Sigma'})\circ \phi Q\]
  \[= \sum_{i\in [n]} p_{i} \interp{Q_{i}, (a_{i}, \dD_{i}, b_{i})} = \sum_{i\in [n]} p_{i}  T( \interp{a_{i}, \dD_{i}, b_{i}} \otimes \interp{\Sigma_{2}}\otimes \interp{\Sigma'} ) \circ \phi Q_{i}\]
  \[ = \sum_{i\in [n]} p_{i}  T( (\boxt \circ p\delta{\curry{\theta_{\dD_{i}}}}) \otimes \interp{\Sigma_{2}}\otimes \interp{\Sigma'} ) \circ \phi Q_{i},\]
  where $\phi Q_{i} : I \to T(\interp{\Sigma_{2}}\otimes \interp{\Sigma'})$, and $\theta_{\dD_{i}} = \interp{b_{i}} \circ \dD_{i}\circ \interp{a_{i}}^{-1}: \interp{S} \to \interp{U}$, and $\dD_{i} : \interp{\Sigma_{a_{i}}} \to \interp{\Sigma_{b_{i}}} $, and
  $\Sigma_{a_{i}} \vdash_{1} a_{i} : S$ and $\Sigma_{b_{i}} \vdash_{1} b_{i} : U$.

  Moreover, by induction hypothesis we also have
  \[ \interp{Q_{i}, N} = \mu \circ T s \circ T(\interp{N}\otimes \interp{\Sigma'})\circ \phi Q_{i} \]
  \[= \sum_{j\in [m]}q_{i,j}\interp{Q_{i,j}', V_{i,j}} = \sum_{j\in [m]}q_{i,j}  T(\interp{V_{i,j}}\otimes \interp{\Sigma'}) \circ \phi Q_{i,j}',\]
  where $\phi Q_{i,j}' : I \to T(\interp{\Sigma_{2}'}\otimes \interp{\Sigma'})$ and $\Sigma_{2}' \vdash_{1} V_{i,j} : S$. 

  Since $\mathrm{operate}(\dD_i, Q_{i,j}', a_{i}, V_{i,j}) = Q_{i,j}''$,
  we have the following in $\q$.
  \[Q_{i,j}'' = ((J\dD_{i} \circ J\interp{a_{i}}^{-1}\circ J\interp{V_{i,j}})\otimes \interp{\Sigma'})\circ Q_{i,j}' : I \to \Sigma_{b_{i}},\Sigma'.\]
  Thus in $\A$, we have
  \[\phi Q_{i,j}'' = \phi((J\dD_{i} \circ J\interp{a_{i}}^{-1}\circ J\interp{V_{i,j}})\otimes \interp{\Sigma'})\circ Q_{i,j}')\]
  \[= \mu \circ T \phi ((J\dD_{i} \circ J\interp{a_{i}}^{-1}\circ J\interp{V_{i,j}})\otimes \interp{\Sigma'})\circ \phi Q_{i,j}' \]
  \[= \mu \circ T (\mu \circ Tt\circ s \circ \phi(J\dD_{i} \circ J\interp{a_{i}}^{-1}\circ J\interp{V_{i,j}})\otimes \eta_{\interp{\Sigma'}})\circ \phi Q_{i,j}'\]
  \[\stackrel{(*)}{=} \mu \circ T (\mu \circ Ts\circ t \circ (\eta_{\interp{\Sigma_{b_{i}}}} \circ \dD_{i} \circ \interp{a_{i}}^{-1}\circ \interp{V_{i,j}})\otimes \eta_{\interp{\Sigma'}})\circ \phi Q_{i,j}'\]
  \[= \mu \circ T (s \circ (\eta_{\interp{\Sigma_{b_{i}}}} \circ \dD_{i} \circ \interp{a_{i}}^{-1}\circ \interp{V_{i,j}})\otimes {\interp{\Sigma'}})\circ \phi Q_{i,j}'\]
  \[=  \mu \circ T (\eta_{\interp{\Sigma_{2}'}\otimes \interp{\Sigma'}} \circ ((\dD_{i} \circ \interp{a_{i}}^{-1}\circ \interp{V_{i,j}})\otimes {\interp{\Sigma'}}))\circ \phi Q_{i,j}'\]
  \[= T((\dD_{i} \circ \interp{a_{i}}^{-1}\circ \interp{V_{i,j}})\otimes {\interp{\Sigma'}}) \circ \phi Q_{i,j}'\]
  
  \[: I \to T(\interp{\Sigma_{b_{i}}}\otimes \interp{\Sigma'}).\]
  Note that $(*)$ is by $\phi \circ J = E \circ \psi$ in condition \ref{embeddings}.
  
  We need to show 
  \[\interp{Q, \mathsf{apply}(M, N)} = \mu \circ Ts \circ T(\interp{\mathsf{apply}(M, N)} \otimes \interp{\Sigma'})\circ \phi Q\]
  \[= \mu \circ Ts \circ T((T\epsilon \circ T((\force \circ \unboxt)\otimes \interp{S}) \circ \mu \circ Ts \circ t \circ (\interp{M}\otimes \interp{N})) \otimes \interp{\Sigma'})\circ \phi Q\]
  \[= \sum_{i,j\in [n]\times [m]}p_{i} q_{i,j} \interp{Q_{i,j}'', b_{i}} =  \sum_{i,j\in [n]\times [m]}p_{i} q_{i,j} (T(\interp{b_{i}}\otimes \interp{\Sigma'})\circ \phi Q_{i,j}'')\]
  \[= \sum_{i,j\in [n]\times [m]}p_{i} q_{i,j} (T(\interp{b_{i}}\otimes \interp{\Sigma'})\circ T((\dD_{i} \circ \interp{a_{i}}^{-1}\circ \interp{V_{i,j}})\otimes {\interp{\Sigma'}})\circ \phi Q_{i,j}')\]
  \[= \sum_{i,j\in [n]\times [m]}p_{i} q_{i,j} (T((\interp{b_{i}}\circ \dD_{i} \circ \interp{a_{i}}^{-1}\circ \interp{V_{i,j}})\otimes \interp{\Sigma'})\circ \phi Q_{i,j}').\]

  This is true because
  \[\mu \circ Ts \circ T((T\epsilon \circ T((\force \circ \unboxt)\otimes \interp{S}) \circ \mu \circ Ts \circ t \circ (\interp{M}\otimes \interp{N})) \otimes \interp{\Sigma'})\circ \phi Q\]
  \[= \mu \circ Ts \circ T((T\epsilon \circ T((\force \circ \unboxt)\otimes \interp{S}) \circ \mu \circ Ts \circ t ) \otimes \interp{\Sigma'} )\circ T((\interp{M}\otimes \interp{N}) \otimes \interp{\Sigma'})\circ \phi Q\]
  \[= \mu \circ Ts \circ T((T\epsilon \circ T((\force \circ \unboxt)\otimes \interp{S}) \circ \mu \circ Ts \circ t ) \otimes \interp{\Sigma'} )\circ T (T\Circ(S, U)\otimes \interp{N}\otimes \interp{\Sigma'} )\]
  \[\circ T(\interp{M}\otimes \interp{\Sigma_{2}} \otimes \interp{\Sigma'})\circ \phi Q\]
  \[\stackrel{(1)}{=} \mu \circ Ts \circ T((T\epsilon \circ T((\force \circ \unboxt)\otimes \interp{S}) \circ t ) \otimes \interp{\Sigma'} )\circ T (\Circ(S, U)\otimes \interp{N}\otimes \interp{\Sigma'})\]
  \[\circ \mu \circ Ts \circ T(\sigma \otimes \interp{\Sigma'}) \circ T(\interp{M}\otimes \interp{\Sigma_{2}} \otimes \interp{\Sigma'})\circ \phi Q\]
  \[\stackrel{IH}{=}\mu \circ Ts \circ T((T\epsilon \circ T((\force \circ \unboxt)\otimes \interp{S}) \circ t ) \otimes \interp{\Sigma'} )\circ T (\Circ(S, U)\otimes \interp{N}\otimes \interp{\Sigma'})\]
  \[\circ \sum_{i\in [n]} p_{i}  (T((\boxt \circ p\delta{\curry{\theta_{\dD_{i}}}}) \otimes \interp{\Sigma_{2}}\otimes \interp{\Sigma'} ) \circ \phi Q_{i})\]
  \[= \sum_{i\in [n]} p_{i} (\mu \circ Ts \circ T((T\epsilon \circ T((\force \circ \unboxt)\otimes \interp{S}) \circ t ) \otimes \interp{\Sigma'} )\circ T (\Circ(S, U)\otimes \interp{N}\otimes \interp{\Sigma'})\]
  \[ \circ T((\boxt \circ p\delta{\curry{\theta_{\dD_{i}}}}) \otimes \interp{\Sigma_{2}}\otimes \interp{\Sigma'} ) \circ \phi Q_{i})\]
  \[= \sum_{i\in [n]} p_{i} (\mu \circ Ts \circ T((T\epsilon \circ T((\force \circ \unboxt)\otimes \interp{S}) \circ t ) \otimes \interp{\Sigma'} )\circ T((\boxt \circ p\delta{\curry{\theta_{\dD_{i}}}}) \otimes T\interp{S}\otimes \interp{\Sigma'}) \]
  \[\circ T(\interp{N}\otimes \interp{\Sigma'}) \circ \phi Q_{i})\]
  \[\stackrel{(2)}{=} \sum_{i\in [n]} p_{i} (T(\epsilon \otimes \interp{\Sigma'}) \circ T ((\force \circ \unboxt) \otimes \interp{\Sigma'})
    \circ T((\boxt \circ p \delta{\curry{\theta_{\dD_{i}}}})
    \otimes \interp{S}\otimes \interp{\Sigma'}) \circ T(\lambda^{-1}\otimes \interp{\Sigma'}) \]
  \[\circ \mu \circ Ts \circ T(\interp{N}\otimes \interp{\Sigma'}) \circ \phi Q_{i})\]

  \[= \sum_{i\in [n]} p_{i} (T(\epsilon \otimes \interp{\Sigma'}) \circ T(\curry{\theta_{\dD_{i}}} \otimes \interp{S}\otimes \interp{\Sigma'}) \circ T(\lambda^{-1}\otimes \interp{\Sigma'}) \circ \sum_{j\in [m]}q_{i,j} (T(\interp{V_{i,j}}\otimes \interp{\Sigma'}) \circ \phi Q_{i,j}'))\]
  \[= \sum_{i, j\in [n]\times [m]} p_{i}q_{i,j} (T(\theta_{\dD_{i}} \otimes \interp{\Sigma'}) \circ T(\interp{V_{i,j}}\otimes \interp{\Sigma'}) \circ \phi Q_{i,j}')\]
  \[= \sum_{i, j\in [n]\times [m]} p_{i}q_{i,j} (T((\interp{b_{i}}\circ \dD_{i} \circ \interp{a_{i}}^{-1})\otimes \interp{\Sigma'}) \circ T(\interp{V_{i,j}}\otimes \interp{\Sigma'}) \circ \phi Q_{i,j}')\]
  \[= \sum_{i, j\in [n]\times [m]} p_{i}q_{i,j} (T((\interp{b_{i}}\circ \dD_{i} \circ \interp{a_{i}}^{-1} \circ \interp{V_{i,j}})\otimes \interp{\Sigma'}) \circ \phi Q_{i,j}').\]
   
  To prove $(1)$, we just need to prove
  \[T((\mu \circ Ts \circ t ) \otimes \interp{\Sigma'} )\circ T (T\Circ(S, U)\otimes \interp{N}\otimes \interp{\Sigma'} )\]
  \[= T((t \otimes \interp{\Sigma'} )\circ T (\Circ(S, U)\otimes \interp{N}\otimes \interp{\Sigma'})\circ \mu \circ Ts \circ T(s \otimes \interp{\Sigma'}).\] 
  This is true because the following diagram commutes.
  \[

\begin{tikzcd}[column sep=-6em, row sep=3em]
  &&& T((T\interp{\Circ(S,U)}\otimes \interp{\Sigma_{2}})\otimes \interp{\Sigma'}) \arrow[dl, "T(s \otimes \interp{\Sigma'})"] \arrow[dr, "T((T\interp{\Circ(S,U)}\otimes \interp{N})\otimes \interp{\Sigma'})"]&& \\
  &&T(T(\interp{\Circ(S, U)}\otimes \interp{\Sigma_{2}})\otimes \interp{\Sigma'})
  \arrow[dl, "Ts"]
  \arrow[dr, "T(T(\interp{\Circ(S, U)}\otimes \interp{N})\otimes \interp{\Sigma'})"]& & T((T\interp{\Circ(S, U)}\otimes T\interp{S})\otimes \interp{\Sigma'})
  \arrow[dl, "T(s \otimes \interp{\Sigma'})"]
  \arrow[dr, "T(t \otimes \interp{\Sigma'})"]& \\
  & TT((\interp{\Circ(S, U)}\otimes \interp{\Sigma_{2}})\otimes \interp{\Sigma'})
  \arrow[dr, "TT((\interp{\Circ(S, U)}\otimes \interp{N})\otimes \interp{\Sigma'})"]
  \arrow[dl, "\mu"]&&T(T(\interp{\Circ(S, U)}\otimes T\interp{S})\otimes \interp{\Sigma'})
  \arrow[dr, "T(T t \otimes \interp{\Sigma'})"]
  \arrow[dl, "Ts"]&& T(T(T\interp{\Circ(S, U)}\otimes \interp{S})\otimes \interp{\Sigma'})
  \arrow[dl, "T(Ts \otimes \interp{\Sigma'})"]\\
  T((\interp{\Circ(S, U)}\otimes \interp{\Sigma_{2}})\otimes \interp{\Sigma'})
  \arrow[dr, "T((\interp{\Circ(S, U)}\otimes \interp{N})\otimes \interp{\Sigma'})"] && TT((\interp{\Circ(S, U)}\otimes T\interp{S})\otimes \interp{\Sigma'})
  \arrow[dl, "\mu"]
  \arrow[dr, "TT(t\otimes \interp{\Sigma'})"]&& T(TT(\interp{\Circ(S, U)}\otimes \interp{S})\otimes \interp{\Sigma'})
  \arrow[dl, "Ts"]
  \arrow[dr, "T(\mu \otimes \interp{\Sigma'})"]& \\
  &T((\interp{\Circ(S, U)}\otimes T\interp{S})\otimes \interp{\Sigma'})
  \arrow[rrrr, bend right = 10, "T(t\otimes \interp{\Sigma'})"]&&TT(T(\interp{\Circ(S, U)}\otimes \interp{S})\otimes \interp{\Sigma'})
  \arrow[rr, "\mu"]& & T(T(\interp{\Circ(S, U)}\otimes \interp{S})\otimes \interp{\Sigma'})
\end{tikzcd}
\]

  To prove $(2)$, we just need to show
  {\small
  \[\mu \circ Ts \circ T((T\epsilon \circ T((\force \circ \unboxt)\otimes \interp{S}) \circ t ) \otimes \interp{\Sigma'} )\circ T((\boxt \circ p\delta{\curry{\theta_{\dD_{i}}}}) \otimes T\interp{S}\otimes \interp{\Sigma'}) \]
  \[= T(\epsilon \otimes \interp{\Sigma'}) \circ T ((\force \circ \unboxt) \otimes \interp{\Sigma'})
    \circ T((\boxt \circ p \delta{\curry{\theta_{\dD_{i}}}})
    \otimes \interp{S}\otimes \interp{\Sigma'}) \circ T(\lambda^{-1}\otimes \interp{\Sigma'})\circ \mu \circ Ts. \]}  

  This is true because the following diagram commutes.
  \[

\begin{tikzcd}[column sep=-4em, row sep=2.5em]
  & & & T(T\interp{S}\otimes \interp{\Sigma'})
  \arrow[rr, "T(\lambda^{-1}\otimes \interp{\Sigma'})"]
  \arrow[dl, "Ts"]
  \arrow[dr, "T(T\lambda^{-1} \otimes \interp{\Sigma'})"]
  & & T(I\otimes T\interp{S}\otimes \interp{\Sigma'})
  \arrow[dl, "T(t\otimes \interp{\Sigma'})"]
  \arrow[dr, "T((\boxt \circ p\delta{\curry{\theta}}) \otimes T\interp{S}\otimes \interp{\Sigma'})"]&&  \\
  & &  TT(\interp{S}\otimes \interp{\Sigma'})
  \arrow[dr, "TT(\lambda^{-1}\otimes \interp{\Sigma'})"]
  \arrow[dl, "\mu"]&
  &T(T(I\otimes \interp{S})\otimes \interp{\Sigma'})
  \arrow[dl, "Ts "]
  &
  &T(\interp{\Circ(S, U)}\otimes T\interp{S}\otimes \interp{\Sigma'})
  \arrow[dl, "T(t\otimes \interp{\Sigma'})"]
  & \\
  & T(\interp{S}\otimes \interp{\Sigma'})
  \arrow[dr, "T(\lambda^{-1}\otimes \interp{\Sigma'})"]& & TT(I \otimes \interp{S}\otimes \interp{\Sigma'})
  \arrow[dl, "\mu"]
  \arrow[dr, "TT((\boxt \circ p \delta{\curry{\theta}}) \otimes \interp{S}\otimes \interp{\Sigma'})"]&&T(T(\interp{\Circ(S, U)}\otimes \interp{S}) \otimes \interp{\Sigma'})
  \arrow[dl, "Ts"]
  \arrow[dr, "T(T((\force \circ \unboxt) \otimes \interp{S}) \otimes \interp{\Sigma'})"]&& \\
  && T(I\otimes \interp{S}\otimes \interp{\Sigma'})
  \arrow[dr, "T((\boxt \circ p \delta{\curry{\theta}}) \otimes \interp{S}\otimes \interp{\Sigma'})", swap]&
  & TT(\interp{\Circ(S, U)} \otimes \interp{S}\otimes \interp{\Sigma'})
  \arrow[dl, "\mu"]
  \arrow[dr, "TT( ((\force \circ \unboxt) \otimes \interp{S})\otimes \interp{\Sigma'})"]&& T(T((\interp{S}\multimap \interp{U})\otimes \interp{S}) \otimes \interp{\Sigma'})
  \arrow[dr, "T(T\epsilon \otimes \interp{\Sigma'})"]
  \arrow[dl, "Ts"]
  & \\
  & && T(\interp{\Circ(S, U)} \otimes \interp{S}\otimes \interp{\Sigma'})
  \arrow[dr, "T((\force \circ \unboxt)\otimes \interp{S}\otimes \interp{\Sigma'})", swap]&& TT(((\interp{S}\multimap \interp{U})\otimes \interp{S}) \otimes \interp{\Sigma'})
  \arrow[dl, "\mu"]&& T(T\interp{U}\otimes \interp{\Sigma'})
  \arrow[dl, "Ts"]\\
  &&& & T((\interp{S}\multimap \interp{U}) \otimes \interp{S}\otimes \interp{\Sigma'})
  \arrow[dr, "T(\epsilon \otimes \interp{\Sigma'})", swap]&& TT(\interp{U}\otimes \interp{\Sigma'}) \arrow[dl, "\mu"]& \\
  &&& && T(\interp{U}\otimes \interp{\Sigma'})&&
\end{tikzcd}
\] 
      \end{itemize}
\end{proof}

\end{document}